\documentclass[longauth]{aa}  
\usepackage[hidelinks,colorlinks=true,linkcolor=blue,citecolor=blue]{hyperref}
\usepackage{graphicx}
\usepackage{txfonts}
\usepackage{array}
\newcolumntype{P}[1]{>{\centering\arraybackslash}p{#1}}
\usepackage{mathtools,cuted}
\usepackage{subcaption}
\usepackage{amsmath}	
\usepackage{amssymb}	
\usepackage{multicol}   
\usepackage{bm}		    
\usepackage{caption}
\usepackage{pdflscape}	
\usepackage[dvipsnames]{xcolor}
\usepackage{lipsum}
\usepackage[utf8]{inputenc}
\newcommand{\hi}{H{\sc i}}

\definecolor{BAROSI}{rgb}{0, 70.2, 70.2}
\newcommand{\alex}{\textcolor{Black}}

\begin{document} 

\title{The BINGO Project III:} \subtitle{Optical design and optimisation of the focal plane}
%
\author{Filipe~B.~Abdalla \inst{1,2,3,4}\fnmsep\thanks{filipe.abdalla@gmail.com},
Alessandro~Marins \inst{2}\fnmsep\thanks{alessandro.marins@usp.br},
Pablo~Motta \inst{2}\fnmsep\thanks{pablo.motta@usp.br},
Elcio~Abdalla \inst{2},
Rafael~M.~Ribeiro\inst{2},
Carlos~A.~Wuensche\inst{3},
Jacques~Delabrouille\inst{5,6,7},
Karin~S.~ F.~Fornazier\inst{2,3},
Vincenzo~Liccardo\inst{3},
Bruno~Maffei\inst{8},
Eduardo~J.~ de~Mericia\inst{3},
Carlos~H.~N.~Otobone\inst{2},
Juliana~F.~ R.~dos~ Santos\inst{2},
Gustavo~B.~Silva\inst{2},
Jordany~Vieira\inst{2},
Jo\~ao A. M. Barretos\inst{2},
Luciano~Barosi\inst{9},
Francisco~A.~Brito\inst{9},
Amilcar~R.~Queiroz\inst{9},
Thyrso~Villela\inst{3,10,11},
Bin~Wang\inst{12,13},
Andre~A.~Costa\inst{12},
Elisa~G.~M.~Ferreira\inst{2,14},
Ricardo~G.~Landim\inst{15},
Camila~Paiva~Novaes\inst{3},
Michael~W.~Peel\inst{16,17}
Larissa~Santos\inst{12,13},
Marcelo~V.~dos~Santos\inst{9},
Jiajun~Zhang\inst{18}}
\institute{Department of Physics and Astronomy, University College London, Gower Street, London,WC1E 6BT, UK 
\and 
Instituto de F\'isica, Universidade de S\~ao Paulo - C.P. 66318, CEP: 05315-970, S\~ao Paulo, Brazil
\and 
Instituto Nacional de Pesquisas Espaciais, Divis\~ao de Astrof\'isica,  Av. dos Astronautas, 1758, 12227-010 - S\~ao Jos\'e dos Campos, SP, Brazil
\and 
Department of Physics and Electronics, Rhodes University, PO Box 94, Grahamstown, 6140, South Africa
\and 
Laboratoire Astroparticule et Cosmologie (APC), CNRS/IN2P3, Universit\'e Paris Diderot, 75205 Paris Cedex 13, France 
\and 
IRFU, CEA, Universit\'e Paris Saclay, 91191 Gif-sur-Yvette, France 
\and 
Department of Astronomy, School of Physical Sciences, University of Science and Technology of China, Hefei, Anhui 230026
\and 
Institut d’Astrophysique Spatiale, Orsay (CNRS-INSU), France
\and 
Unidade Acad\^emica de F\'{i}sica, Univ. Federal de Campina Grande, R. Apr\'{i}gio Veloso, 58429-900 - Campina Grande, Brazil
\and 
Instituto de F\'{i}sica, Universidade de Bras\'{i}lia, Bras\'{i}lia, DF, Brazil 
\and 
Centro de Gest\~ao e Estudos Estrat\'egicos - CGEE,
SCS Quadra 9, Lote C, Torre C S/N Salas 401 - 405, 70308-200 - Bras\'ilia, DF, Brazil
\and 
 Center for Gravitation and Cosmology, College of Physical Science and Technology, Yangzhou University, 225009, China
\and 
School of Aeronautics and Astronautics, Shanghai Jiao Tong University, Shanghai 200240, China 
\and 
 Max-Planck-Institut f{\"u}r Astrophysik, Karl-Schwarzschild Str. 1, 85741 Garching, Germany
\and 
 Technische Universit\"at M\"unchen, Physik-Department T70, James-Franck-Stra\text{$\beta$}e 1, 85748 Garching, Germany
\and 
Instituto de Astrof\'{i}sica de Canarias, 38200, La Laguna, Tenerife, Canary Islands, Spain 
\and 
Departamento de Astrof\'{i}sica, Universidad de La Laguna (ULL), 38206, La Laguna, Tenerife, Spain
\and 
Center for Theoretical Physics of the Universe, Institute for Basic Science (IBS), Daejeon 34126, Korea
}
\authorrunning{Abdalla, Marins, Motta et al.}
\date{Received MMM XX, YYYY; accepted MMM XX, YYYY}
\titlerunning{The BINGO Project III: Optical design and optimisation of the focal plane} 
\abstract
{The BINGO telescope was designed to measure the fluctuations of the 21-cm radiation arising from the hyperfine transition of neutral hydrogen and aims to measure the Baryon Acoustic Oscillations (BAO) from such fluctuations, therefore serving as a pathfinder to future deeper intensity mapping surveys. The requirements for the Phase 1 of the projects consider a large reflector system (two 40 m-class dishes in a crossed-Dragone configuration), illuminating a focal plane with 28 horns to measure the sky with two circular polarisations in a drift scan mode to produce measurements of the radiation in intensity ($I$) as well as the circular ($V$) polarisation.}
{In this paper we present the optical design for the instrument.  We describe the optical arrangement of the horns in the focal plane to produce a homogeneous and well-sampled map after the end of Phase 1, as well as the intensity and polarisation properties of the beams. Our analysis provides an optimal model for the location of the horns in the focal plane, producing a homogeneous and Nyquist sampled map after the nominal survey time.}
{We used the GRASP package to model the focal plane arrangement and performed several optimisation tasks to arrive to the current  configuration, including an estimation of the side lobes corresponding to the diffraction patterns of the two mirrors. The final model for the focal plane was defined through a combination of neural network and other direct optimisation methods.}
{We arrive at an optimal configuration for the optical system, including the focal plane positioning and the beam behavior of the instrument. We present an estimate of the expected side lobes both for intensity and polarisation, as well as the effect of band averaging on the final side lobes, as well as an estimation of the cross-polarization leakage for the final configuration.}
{We conclude that the chosen optical design meets the requirements for the project in terms of polarisation purity, area coverage as well as homogeneity of coverage so that BINGO can perform a successful BAO experiment. We further conclude that the requirements on the placement and r.m.s. error on the mirrors are also achievable so that a successful experiment can be conducted.}
\keywords{optical design -- astrophysics 21-cm --  cosmology  -- Baryon Acoustic Oscillations}
\maketitle
%
%
\section{Introduction}
%
The BINGO (Baryon Acoustic Oscillations [BAO] from Integrated Neutral Gas Observations) project is a two dish radio telescope\footnote{Although we have two dishes, the telescope works as a single dish radiometer, see paper II of this series.} aiming to observe the 21-cm line corresponding to the hyperfine transition of neutral atomic Hydrogen (\hi).  It will survey a sky area of  $\sim 5324$ square degrees, i.e. about 13\% of the sky, in a redshift range spanning from 0.127 to 0.449 (corresponding to a frequency span of 980\,MHz to 1260\,MHz). 

Large scale structure maps of the Universe have been performed with optical galaxy spectroscopic surveys \citep{2005ApJ...633..560E,2011MNRAS.416.3017B,2011MNRAS.415.2892B,2020MNRAS.tmp.2651B,2020ApJ...901..153D}, however only a small fraction of the Universe has been mapped with such techniques. Obviously, the advent of multi object spectrographs can help the optical cosmologist to pave this way with greater speed \citep{2019BAAS...51g..57L}. When we observe the sky in the radio part of the spectrum, we are able to map such spectroscopic information of the 21-cm line without having to specifically obtain the redshift of individual objects because we obtain information at each frequency for the entire field. We obviously cannot identify each galaxy individually if the spatial resolution prevents us from doing so, however can directly obtain information about the redshifts of the galaxies present in the beam simply because there is little confusion between the 21-cm line and other emission in the radio part of the spectrum. 

It has been proposed that mapping the Universe measuring the collective 21-cm line emission of the underlying matter \citep[e.g.,][]{1997ApJ...475..429M,2004MNRAS.355.1339B}, at low resolution, can be much more efficient than doing so at higher resolutions or with optical spectrographs. Such a technique is called Intensity Mapping \citep[IM;][]{2009astro2010S.234P}. The IM observation technique uses, in its most common realization, the 21-cm line emission line of HI, but possibilities of exercising this technique also exist using other lines, such as carbon monoxide lines \citep[e.g.,][]{2011ApJ...741...70L,2012ApJ...745...49G,2014ApJ...786..111P}.  \hi\,21-cm line IM has been proposed as the main technique in several other projects such as the MeerKAT \citep{2017arXiv170906099S}, Tianlai \citep{2012IJMPS..12..256C}, CHIME \citep{2014SPIE.9145E..22B}, HIRAX \citep{2016SPIE.9906E..5XN}, HERA \citep{2017PASP..129d5001D} as well as the SKA \citep{2020PASA...37....7S}. The BINGO project will be an excellent experiment to test IM performance in our redshift range. Details of IM and component separation approaches are described in two companion papers \citep{2020_sky_simulation,2020_component_separation}.


The BINGO telescope will measure \hi\,in emission, the most common component present in our Universe. If we assume that the \hi\,distribution follows the matter distribution in the Universe, we will be in a position to have a detailed map of the matter in intensity maps \citep{2013MNRAS.434L..46S}.   
This will allow us to be the first experiment to measure the BAO in the radio frequency band, although previous BAO detections have already been achieved in optical surveys \citep[e.g.,][]{2005ApJ...633..560E, GBTcross2013, GBTcross2015, GBTcross2021}. 

The desired BAO signal depends on the width of the chosen bins in redshift, and for binnings which correspond to around 0.05 in redshift (for an optimisation of the binning width see \citealt{2020_forecast}), this signal is of the order of a few hundred $\mu $K, more than  1000 times weaker than the smooth spectrum foreground signal. The relative strength of the foregrounds, which are partially linearly polarised and concentrated towards the Galactic plane, means that the observations need to be made with a clean beam, and minimize RFI pick-up \citep{1f_HI_2018,Peel:2019},  with low side-lobe levels and very good polarisation purity. 

The need to resolve structures of angular sizes corresponding to a linear scale of around 150 Mpc in our chosen redshift range implies that the required angular resolution should be close to $\sim 40'$  \citep{2013MNRAS.434.1239B}. The above requirements set out the primary scientific constraints to BINGO's optical design. The aim of this paper is to describe the steps that generated such a design, satisfying the requirements of the project, in terms of polarisation purity, beam shape, homogeneity of the covered area and good sky sampling.


During Phase 1, the instrument should operate with 28 horns, a system temperature $T_{\rm sys} \approx 70$ K per receiver and should be able to measure $I$ and $V$ polarisation \citep{2020_BINGO_Instrument}. The motivation of this paper is to 
describe the development of BINGO optical system relative to Phase 1. The BINGO project is described in \cite{2020_project} and the current status of the project is presented in the companion papers  \citep{2020_project,2020_BINGO_Instrument,2020_sky_simulation,2020_component_separation,2020_mock_simulations,2020_forecast}. 

This paper describes the steps to produce the optical design that meets the BINGO scientific requirements. Section \ref{Sec:BingoDesign} contains a description of the optical design. Section \ref{sec:focal_plane_arrang} contains a study of the focal plane and description of the possible arrangements of horns in the focal plane. Section \ref{sec:optimasation} sets out an optimisation of the horn responses given their locations in the focal plane. Section \ref{sec:results} outlines the resulting beam profiles, projected beams for the scanning strategy, polarised responses as well as other properties of the optics chosen after optimisation of the horn locations. We conclude our findings in Section \ref{sec:conclusions}.  

\section{BINGO Optical Design}
\label{Sec:BingoDesign}

\subsection{ Science requirements summary}

Before embarking into an outline of the optical model definition, we summarise here the scientific requirements that are currently in the literature. We will discuss these requirements to a large extent throughout the manuscript, however we set out an initial list here of the requirements already accepted in previous analysis \citep{2013MNRAS.434.1239B}. These requirements are summarized in Table \ref{tab:requirement}.

\begin{table}[h]
\footnotesize
\centering
\caption{Requirements presented in \cite{2013MNRAS.434.1239B} for the BINGO telescope.}
\begin{tabular}{c l c }
 \hline
 & Requirement & Value \\
 \hline
(i) & Angular resolution  & 40 arcmin\\ 
(ii) & Operating frequency      & $960-1260$ MHz \\
\alex{(iii)} & Frequency resolution & $\gtrsim 1$ MHz \\ 
\alex{(iv)} & Frequency baseline  as free as & \\  &possible from instrumental ripples  & \\
\alex{(v)}  & Exceptionally stable receivers & \\ 
\alex{(vi)}  & Sky coverage    & $>2000$ deg$^2$ \\
\alex{(vii)} & Number of feeds   & $>50$  \\
(viii)  & Sidelobe levels as low as possible  &   \\
(ix)  &  Beam ellipticity           &  $<0.1$\\
\alex{(x)}   & No moving parts    &     \\
(xi)  &  Projected aperture diameter &   40 m\\ & of the primary reflector    &\\
(xii) & Minimum number of pixels in the  & 50 \\ & focal plane & \\ & Each of them satisfying:   & \\
(xii) -- (a)    &   Forward gain-loss in &  $<1$ dB\\ & comparison to central pixel & \\
(xii) -- (b)  & Cross-polarization better than &  $-30$ dB \\
(xii) -- (c) & Beam ellipticity & $<0.1$\\
\alex{(xiii)} & Focal length & 90 m  \\
\alex{(xiv)} & Instantaneous field of view & $10^\circ \times 9^\circ$\\
\hline
\end{tabular}
\label{tab:requirement}
\end{table}

 The operating frequency range is chosen to avoid strong RFI from mobile phone downlinks. The current frequency interval is $980 - 1260$ MHz and was chosen after the 2017 RFI campaign described in \cite{Peel:2019}. Stable receivers are a must to reach the required sensitivity of $\sim 100$ $\mu$K per pixel, needed for BAO detection, and to minimize the nocive $1/f$ contribution to the system temperature, estimated to be $T_{sys} = 70 K$. These and other instrument requirements are presented in companion paper by  \cite{2020_BINGO_Instrument}.
 
As explained in \cite{2013MNRAS.434.1239B}, 
50 horns were shown to be an optimal number to achieve a good sensitivity for BAO detection. The setup discussed in this paper contains 28 horns in a different focal plane arrangement. This arrangement was tested in mission simulations and deliver the required sensitivity, as discussed in the companion paper by \cite{2020_sky_simulation}. 
Ultimately 28 horns will yield a less sensitive survey than the original 50 horns, however the simulations in this paper and the ones in \cite{2020_sky_simulation} show that this setup provides good enough results for the science to be achieved.

The low sidelobe requirement mentioned in \citep{2013MNRAS.434.1239B} should be achieved, since the measurements of the first sidelobes and beam shape of the prototype horn are well under $-20$ dBi and, in most of the frequency range, below $-30$ dBi. The beam symmetry is also well characterized in those measurements, as described in \cite{2020ExA....50..125W}. These low sidelobes are also needed to avoid ground pickup and RFI contamination from surface sources (mobile stations, ground radio links, etc).

The beam ellipticity shown in Table \ref{tab:requirement} would  allow map-making and
power spectrum analysis routines to work efficiently. A telescope with no moving parts would be stable, simple and  low-cost. An   underilluminated primary reflector of 40 m is required in order to reach the resolution better than 40 arcmin at 1 GHz. The other requirements were reached after an extensive analysis presented in sections 3 and 5 of  \cite{2013MNRAS.434.1239B}, and the values presented in Table \ref{tab:requirement} were the baseline for this work.

\subsection{Optical model definition}

The current optical design for BINGO is based on a two mirror, off-axis, crossed Dragonian design \citep{1978ATTTJ..57.2663D,2008ApOpt..47..103T}, which is very common in CMB experiments because it minimizes cross polarization (e.g., \citealp{QUIJOTE2018}), so that the focal plane is at right angles to the incoming light, preserving the polarisation of light. The primary mirror is a parabolic mirror with a  focal length ($f$) of 140 m; the mirror is symmetric around its $z$ axis and is a cut out of this parabola in the ($x$,$y$) plane centered at 226.54 m away from the origin with a diameter ($D$) of 40 m in this same ($x,y$) plane. This reference system will be called the global reference frame and the center of the primary is located at coordinates ($x=-226.54, y=0, z=91.64$) m in this frame. 

The secondary mirror is a hyperbolic mirror. The origin of this hyperbola is placed at the focus point of the parabola described in the last paragraph. Therefore, the origin of this hyperbola reference frame is localized at (0, 0, 140) m of the global reference frame and is rotated relatively to the global system by the following Euler angles: ($180^{\circ}, 95^{\circ}, 0$). 

\begin{figure}[h]
    \centering
    \includegraphics[width=9cm]{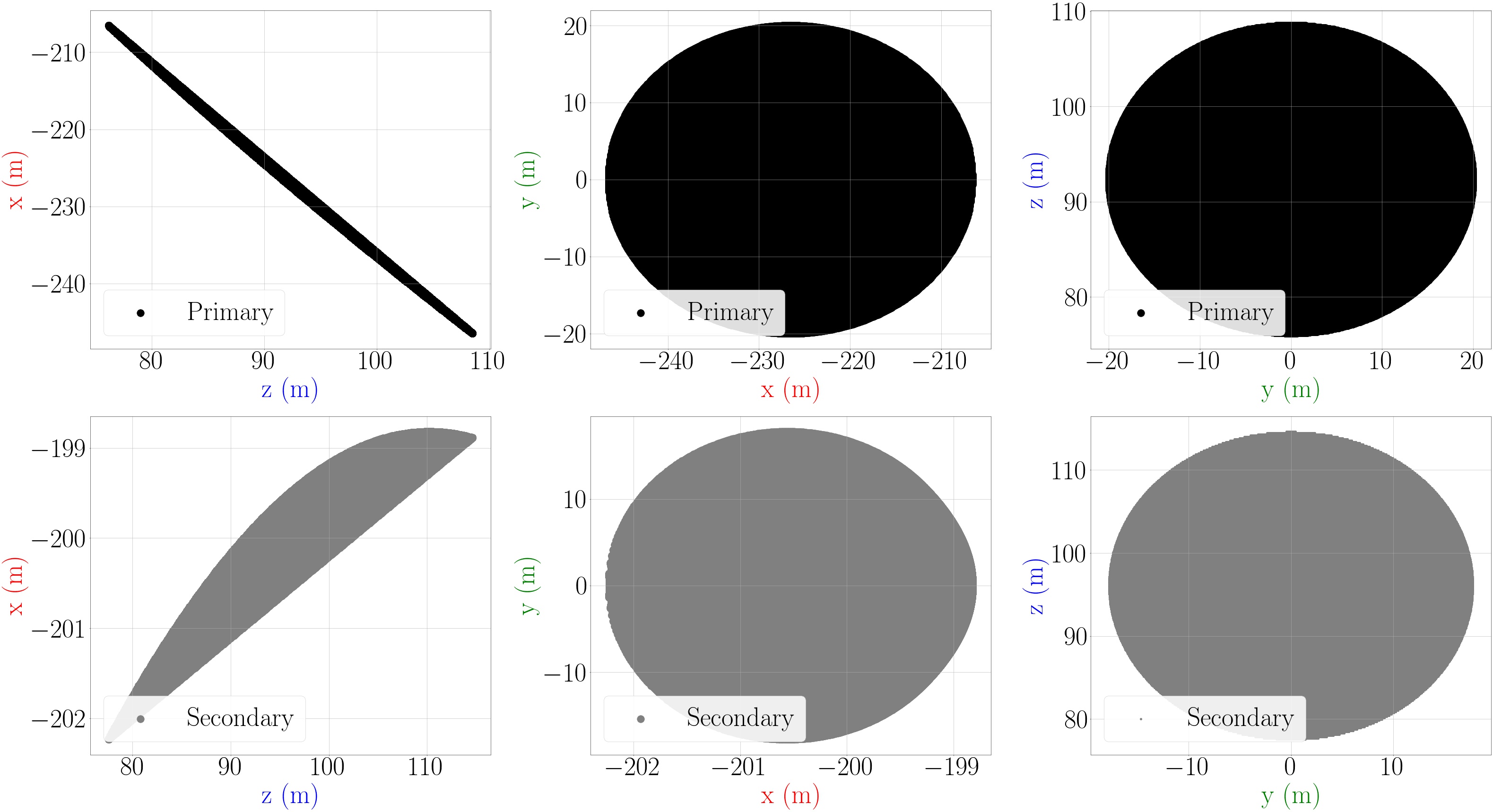}\\
    \includegraphics[width=9cm]{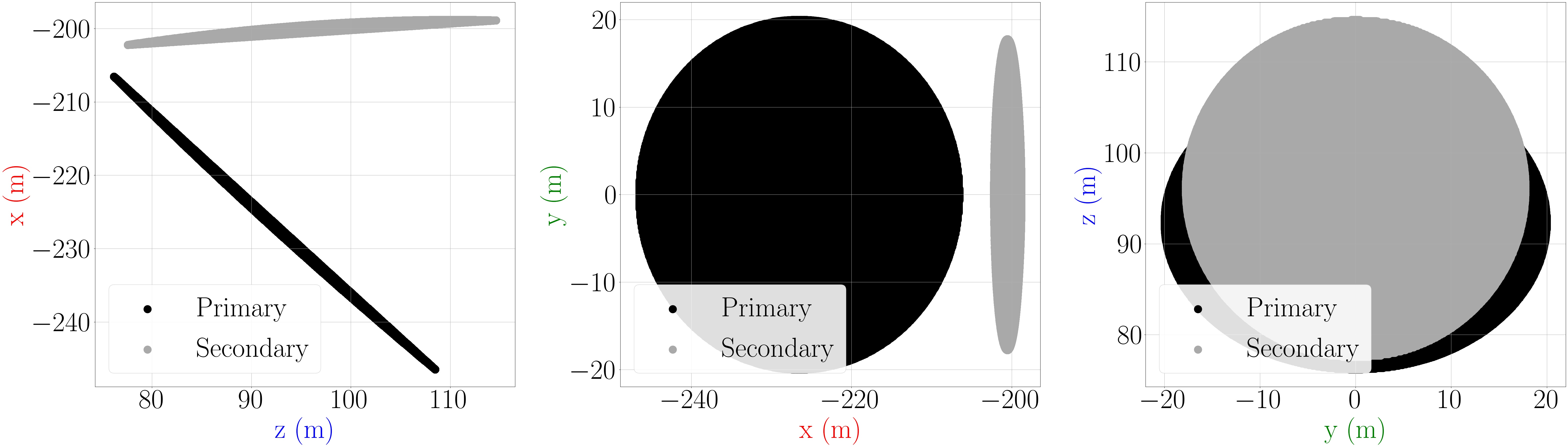}
    \caption{
    Main and sub reflectors representations, in the global coordinate system and 2D-projections. 
    Top row: Primary reflector. Left: Projection in the $(x,z)$ plane. Center: Projection in the $(x,y)$ plane. Right: Projection in the $(z,y)$ plane. Center row: Secondary reflector. Left, center and right are the same as top row. Bottom row: Superposition of primary and secondary reflectors. Left, center and right are the same as top row. }

    \label{fig:reflectors}
\end{figure}

\begin{figure*}[t]
    \centering
    \includegraphics[width=16.0cm,trim=2cm 2cm 2cm 2cm]{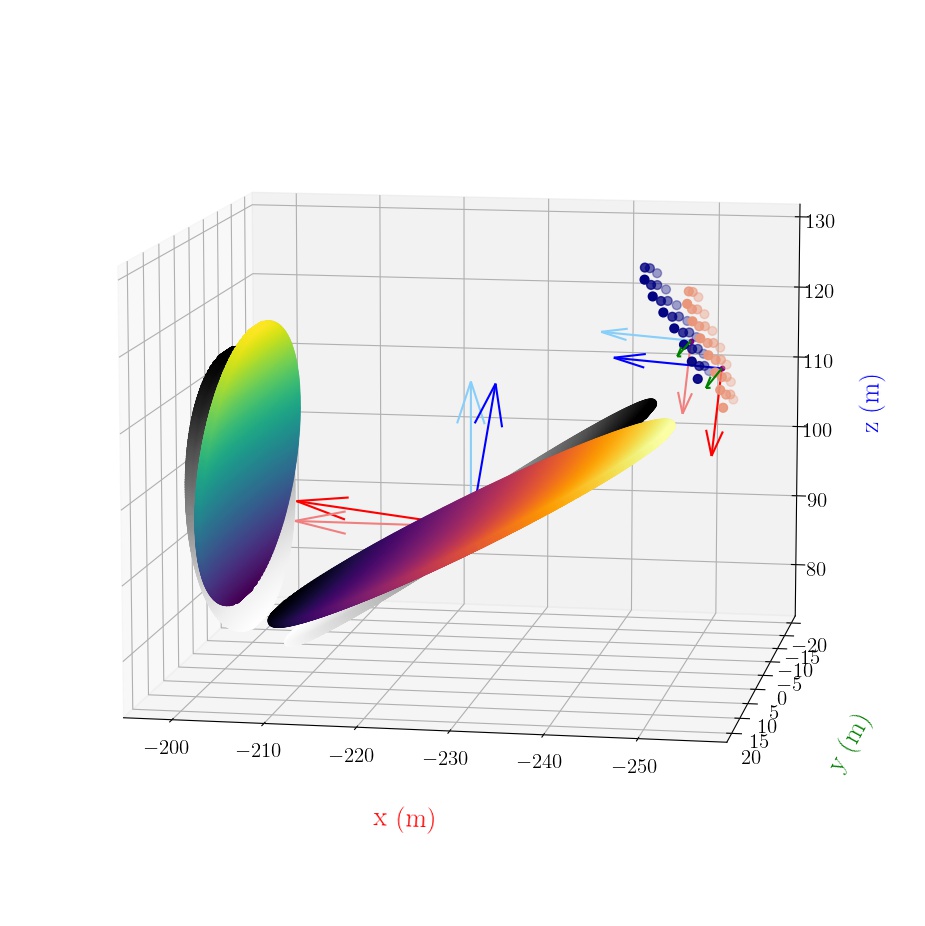}
    \caption{3D plotting of the optical system. The primary reflector in the center, the secondary reflector is on the left and the horn array is on the right. The system is aligned in the North-South direction, with the secondary to the South and the horns to the North. The positions of the mirrors outlined in gray/black are in the global reference system; the colored position reflect the $7^{\circ}$ rotation to account for the pointing of the telescope as described in the text. Horns plotted in salmon colors are in the optimal position.}
    \label{fig:grasp_python}
\end{figure*}

This implies that, for the primary reflector, $f/D = 3.5$ and the angle between the main reflector axis and the secondary reflector axis is $85^{\circ}$. The ratio of the distances between the foci relative to the main reflector focal length is $1.8$ and the sub reflector eccentricity is $-1.7$. This yields an effective focal length for the optical system of $63.20$ m. This value is somewhat smaller than the baseline design discussed in the last session, however this difference makes for a more compact design which is beneficial given that the size of the focal plane is already relatively large. Fig. \ref{fig:reflectors} shows the projections of the primary and secondary dishes onto the global reference frame, where the black color represents primary reflector and gray color represents the secondary reflector with its values in relation to global system. We do not include in this figure the origins and foci of the parabola and the hyperbola given that this would prevent us from looking at the curvature of the mirrors when zoomed in this representation.



Finally, the last coordinate system we set out here defines the location of the focus of the entire system and hence the location of the central feedhorn. This feed reference frame is located at position (0, 0, 252) m of the secondary reference frame (the one defined to locate the hyperbolic mirror) and has a rotation of the following Euler angles: (0, 153.27$^{\circ}$, 180$^{\circ}$). This reference frame will be used later on when we optimise the location of each feed horn in our focal plane. 




In the global reference frame, the optical system is designed to look towards the $z$ direction, therefore rotating the $z$ axis of the global reference frame to coincide with the vertical axis of the site will make the telescope point to the zenith. Since the site coordinates are (Lat: 7$^{\circ}$\, 2'\, 27.6''\, S; Lon: 38$^{\circ}$\, 16'\,  4.8''\, W), rotating the optical system by $7.95^{\circ}$ south (negative direction of the global system $x$ axis), will point the the central horn of the optical system to $\delta=-15^{\circ}$. This configuration is plotted in Fig. \ref{fig:grasp_python}.

We implement the above optical configuration in the
GRASP package (TICRA - Reflector Antenna and EM Modelling Software)\footnote{\url{https://www.ticra.com/software/grasp/}} software 
to investigate four options of horn positioning in the focal plane to search for an optimal illumination and beam forming.

\subsection{Optical arrangement implications for horn requirements\label{section:spillover}}

The theoretical design of the horns expect a return loss of $\sim -40$\,dB across the band. A horn prototype has already been designed and fully tested. The results were reported by \cite{2020ExA....50..125W} and the parameters described there are adopted for this study. 

Together with the low sidelobe level (below $-27$\,dB  below peak for the first sidelobe), we require a spillover below 2$\%$ in order to satisfy our science requirements. 

GRASP builds the simulation with the rays from horn to secondary reflector, and the rays reflected by it in direction of the main reflector. Therefore, for each reflector there will be an associated spillover. We denote 'spillover1' for the secondary reflector and 'spillover2' for the main reflector. 

Since the field (in relation to the rays) contains the power expressed by Poynting's vector $\vec{S}$ $=\frac{1}{2} \mathrm{Re}(\vec{E} \times \vec{H}^{\dagger}$), the power hitting a surface is
$$
W = - \int_{\mathcal{S}}\mathrm{d}\vec{a}\cdot\vec{S},
$$

\noindent with $\mathcal{S}$ reflector surface. So, the spillover in the software is defined as

$$
spillover = 10\log_{10}\frac{4\pi}{W}.
$$ 

We use GRASP to model the optical configuration in later sections and derive an estimate for the spillover (see section \ref{section:spillover} for more comments) throughout the focal plane.
These specifications determine the shape of the input response of the horn that was used in this paper and which is described in this next sub-section. 

\subsection{Signal attenuation from horn design}
\label{sec:taper}

The horns need to have very low side-lobes and the beam needs to be tapered so the illumination at the edge of the secondary mirror is less than $-20$\, compared to the center, minimizing the spillover and ground pickup contributions to the incoming signal.

The resulting beam pattern is important to allow the faint \hi\, signal to be efficiently separated from the bright Galactic foreground emission, whose signal is about 4 orders of magnitude brighter than \hi. 

In the optical system presented in Fig. \ref{fig:grasp_python}, the half-angle subtended by the sub-reflector at the feed center is 17.98$^{\circ}$. With slight under-illumination of the secondary, the resulting full width at half maximum (FWHM) for the whole telescope should be $\approx$ 40 arc minutes (at the central frequency of our band), maintaining the original angular resolution requirements for the project. 

This is mainly determined by the diffraction limit of the primary mirror but will of course depend on the details of the location of the horns and potential optical aberrations present in the system for horns which are not located at the exact focus of the telescope. Each horn in the focal plane will have a different full width half maximum (FWHM), which will be modelled and shown in the section \ref{sec:results}. We also include images of the optical aberrations present in our focal plane given our configuration of choice.

\begin{figure}[h]
\begin{minipage}[c]{0.5\textwidth}
    \centering
    \includegraphics[width=8cm]{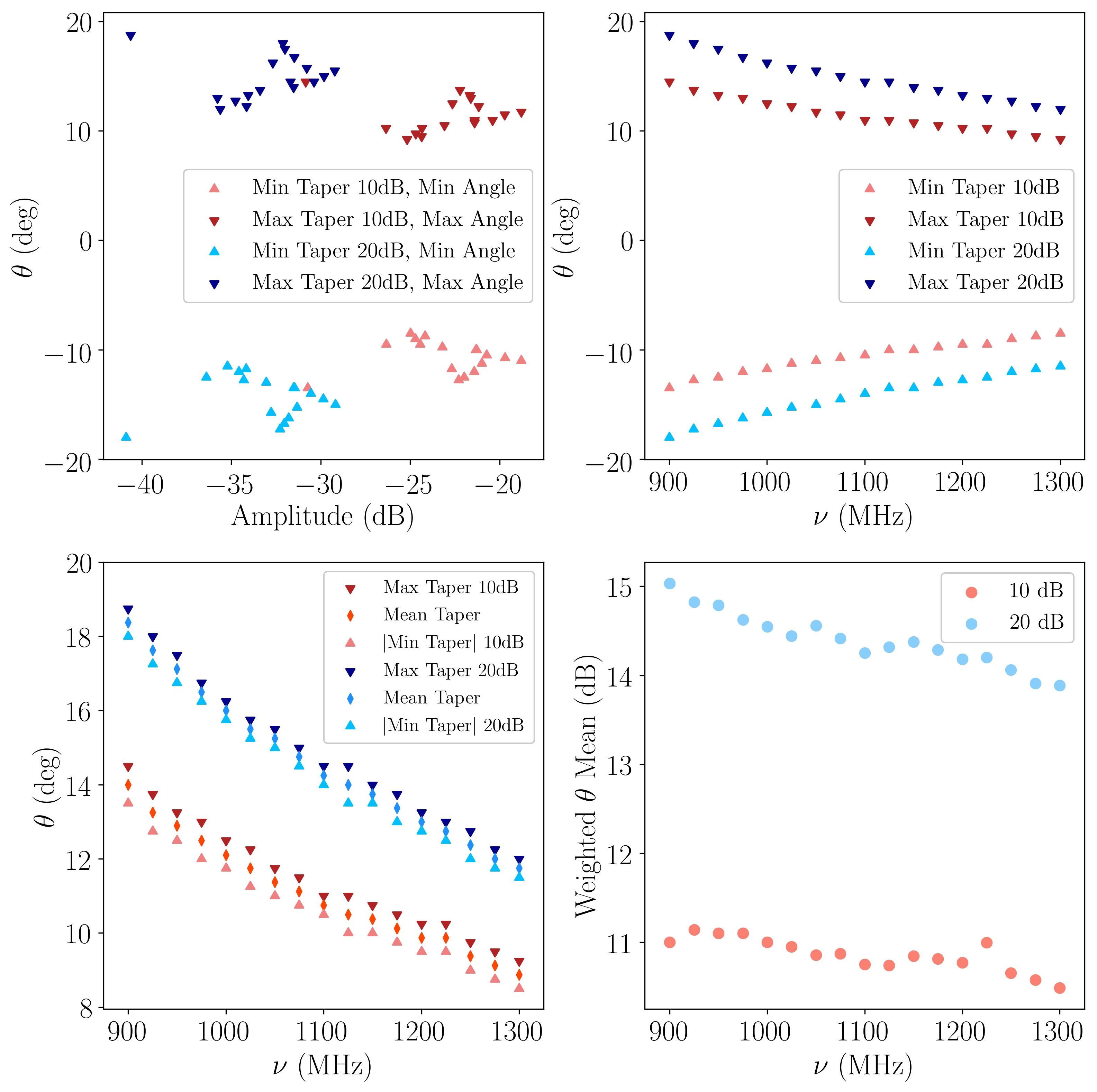}
    \caption{Horizontal polarisation intensity fits for the measurements of the prototype horn. Upper left: \textcolor{black}{10-20 dB attenuation angles values in different frequencies. Upper right: 10-20 dB attenuation angles $\times$ frequency. Lower left: attenuation angles absolute values. Lower right: 10-20 dB weighted attenuation angles $\times$  frequency.} Minimum and maximum values are quoted in relation to maximum beam intensity for each frequency.}
    \label{fig:LIT2018_horiz_results}
\end{minipage}
\begin{minipage}[c]{0.5\textwidth}
\centering
    \includegraphics[width=8cm]{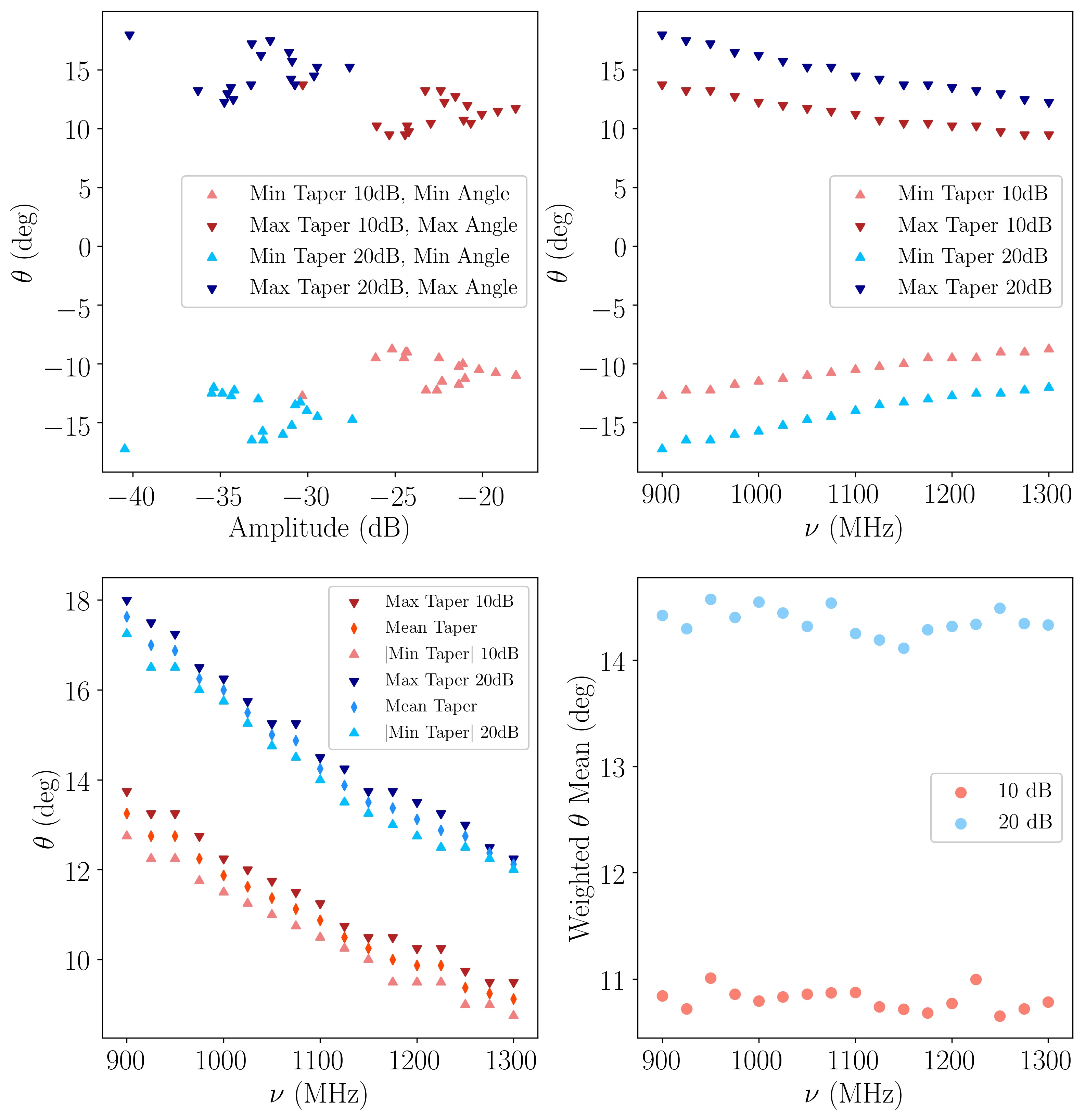}
    \caption{Vertical polarisation intensity fits for the measurements of the prototype horn. Legends are the same as for the horizontal measurements.}
    \label{fig:LIT2018_vert_results}
\end{minipage}
\end{figure}

In the analysis presented in this paper, we used the measurements of the horn radiation patterns reported in \cite{2020ExA....50..125W} to produce fits to the measured horizontal ($H$) and vertical ($V$) polarisation intensities used to model the full beam pattern of the telescope. The fits are show in Appendix \ref{AppendixB}. 
We compute the angle, and corresponding intensity values, for each frequency and each measured ($V$, $H$) intensity component, for which the intensity is attenuated by 10\,dB and 20\,dB in relation to the peak intensity. The results are show in Figures \ref{fig:LIT2018_horiz_results} and  \ref{fig:LIT2018_vert_results} for the horizontal and vertical components, respectively. \textcolor{black}{We stress here that we have modelled the horns as Gaussian beams following the fits outlined above, as this are very good fits to the measurements in \cite{2020ExA....50..125W}, furthermore, we have neglected any cross coupling of the horns in this work. We also assume that the GRASP Gaussian beam is placed at the phase center of the physical horn which is locate 30 cm inwards from the mouth of the horn.} 

Fig. \ref{fig:LIT2018_horiz_results} contains four plots: upper left show the angles for which the intensity is attenuated by 10 and 20\,dB in relation to peak intensity (\textcolor{black}{hereafter, we will refer to these simply by \emph{10} and/or \emph{20 dB attenuation angles}}.); upper right shows the \textcolor{black}{10 and 20 dB attenuation angles} as function of the frequency---note the clear reduction of the angle opening as the frequency increases; lower right shows the maximum, mean and minimum attenuation angles as a function of frequency, for 10 and 20\,dB; lower right shows the weighted average ($\theta \times \nu/(1000\ \textrm{MHz})$ of the attenuation angles as a function of frequency. The same discussion applies to Fig. \ref{fig:LIT2018_vert_results}.

\begin{figure}
\centering
\includegraphics[width=8.5cm]{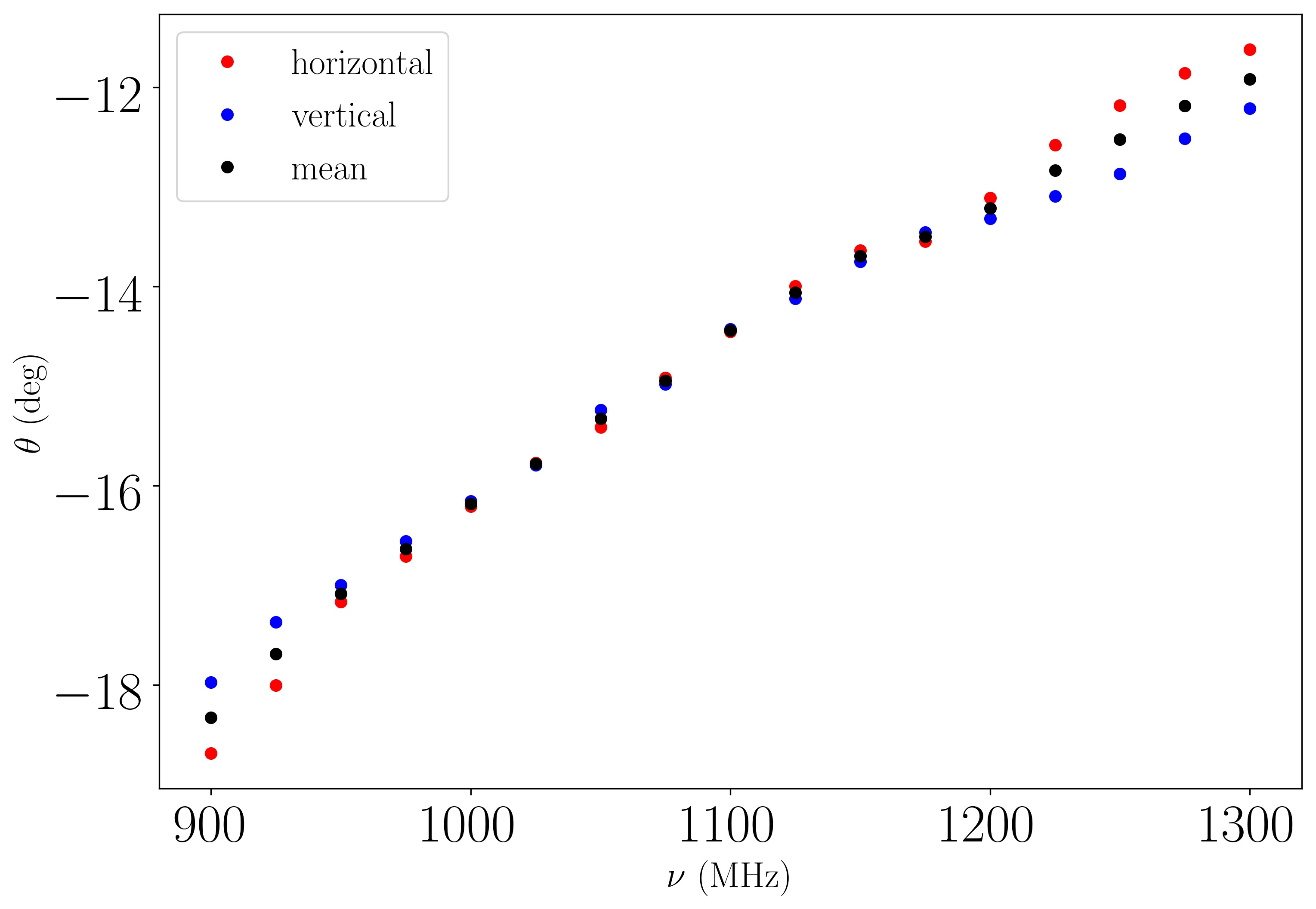}
\caption{The $\theta$ attenuation angles values for vertical (blue) and horizontal (red) polarisation intensity component in different frequencies and mean (black) value between both components in the same frequency.}
\label{fig:LIT2008_interpolation}
\end{figure}

We use the \textcolor{black}{20 dB attenuation angles} for both $V$ and $H$ components to compute mean values for each frequency and use the interpolations in this range in the GRASP simulations. The values shown in Fig. \ref{fig:LIT2008_interpolation} are indeed smaller than the angle subtended by the sub-reflector at the feed center (17.98$^{\circ}$) inside the 980--1260\,MHz band, hence we confirm that our optical system is in fact under illuminated, producing the very low sidelobes as per the scientific requirements, as will be seen in later sections.

\section{The focal plane arrangements}\label{sec:focal_plane_arrang}

\begin{figure*}
\centering
\includegraphics[scale=0.45]{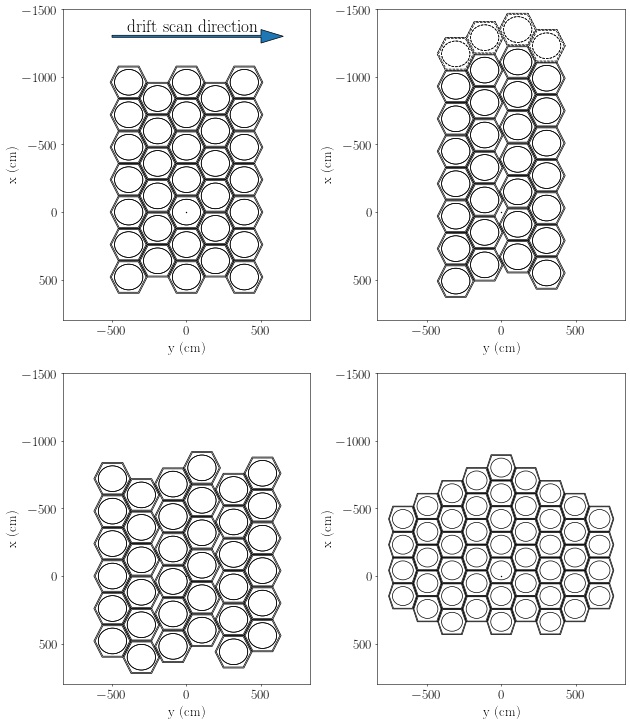}
\caption{The four focal plane arrangements studied in this paper, with different number of horns and different positions in the arrangement. Top left: \emph{Rectangular}; top right: \emph{Double Rectangular}; bottom left: \emph{Triple Rectangular}; bottom right: \emph{Hexagonal}. In the three first arrangements the hexagonal distribution (Structure where the horn is.) has 249.44 cm as maximum width and 240 cm as height. In this structure, the circle represent central position of the horn with 190 cm in diameter. The arrangement four was to build with different measures of the Hexagonal structures to contain the horns, with 192 cm maximum width and 190 cm height. \textcolor{black}{The arrow indicates the BINGO scanning direction in the sky.}} \label{fig:FocalPlaneDesign}
\end{figure*}

The focal plane configurations studied for BINGO are shown in Fig. \ref{fig:FocalPlaneDesign}. For each configuration, we considered the following requirements: a) homogeneous coverage of the sky; b) sampling, in terms of elements of resolution per pixel, of the final maps (to be as close to Nyquist sampling, or better, if possible) and;  c) efficiency of the horns when placed at different locations of the focal plane.

Some original square configurations were not considered here. The collaboration considered at some point a baseline configuration shown on the bottom-right arrangement shown in  Fig. \ref{fig:FocalPlaneDesign}, hereafter the hexagonal configuration). In this configuration the horns are positioned in columns with different number of elements in relation to the central column, where a ``central horn'', which points at $\delta = -15^{\circ}$, is located. 

The idea behind that original configuration is that there is least amount of optical aberrations in the final beams so that the beams are as clean and as close to Gaussian as possible (given that the horns are closely packed and as close to the focus as possible). The configuration is not a complete hexagon because the lowest horns would be obstructed by the primary mirror of the telescope and therefore cannot be considered. In this hexagonal positioning, the horns are encapsulated inside a regular hexagonal case. 

However, we do have to consider that BINGO is a transit telescope  
with no moving reflectors. Therefore, as the sky drifts across the focal plane, it does so across the $y$ direction of the focal plane, where the $y$ coordinate has been taken to be the coordinate in the horn reference system, as plotted in Fig. \ref{fig:FocalPlaneDesign}. Therefore, each horn at a given $x$ location would see the sky at a given declination. If we look carefully at the hexagonal configuration, there are several horns that would see the sky at the same height, therefore at the same declination. This, in theory, would yield an uneven coverage of the sky (which is confirmed by our simulations in a latter section) which is less desirable when it comes to a power spectrum analysis of the underlying residual fields which the 21-cm radiation will leave as an imprint.

The second issue with this hexagonal configuration comes about when we examine the number of pixels per resolution element for such horns. If we want to have a declination range of 15$^{\circ}$ in the sky, and also given that the focal length of the telescope is 63.2 m as discussed previously and the fact that the horns have a diameter of 1.9 m, that means that if the horns are closely packed, the beams would be separated by that ratio which in degrees is $1.73^{\circ}$. Therefore even if we closely pack the horns in a vertical direction we would in fact undersample the focal plane of such an optical system.

If we want a $15^{\circ}$ declination range, which comes from science requirements, and if horns of physical opening 1.9 m are separated in angular size by 1.73$^{\circ}$, it means that our focal plane must be 16.49 m tall. We must note that the hexagonal distribution considered above would be just shy of this height. Therefore we started investigating other distributions that might deliver a better coverage in the desired declination range. If we now calculate the vertical separation needed for the beams to be separated by half a beam width, i.e., having a number of samples per beam equal to 2, therefore being Nyquist sampled, we need to distribute the horns vertically each 37 cm, i.e., the 16.49 m divided by the number of resolution elements inside these 15$^{\circ}$, i.e., 15 divided by the beam size of two thirds of a degree times 2 to account for Nyquist sampling.

The consequence of this is that in the hexagonal structure proposed as the fourth plot in Fig. \ref{fig:FocalPlaneDesign}, several horns would sweep the sky at the exact same declination (this would happen with horns which have same x coordinate). Assuming that the cells that hold the horns are hexagons of height 1.9 m, the adjacent horns would sweep in between gaps where the vertical separation would be 95 cm. This is a sampling which is significantly away from a Nyquist sampling of the focal plane calculated in the previous paragraph. We note here that one possible solution to this problem is to rotate the focal plane of the telescope appropriately, however this solution was discarded as it would increase the complexity of the structure that would be necessary to support the focal plane.

\begin{figure}[!ht]
\centering
\includegraphics[width=9.0cm]{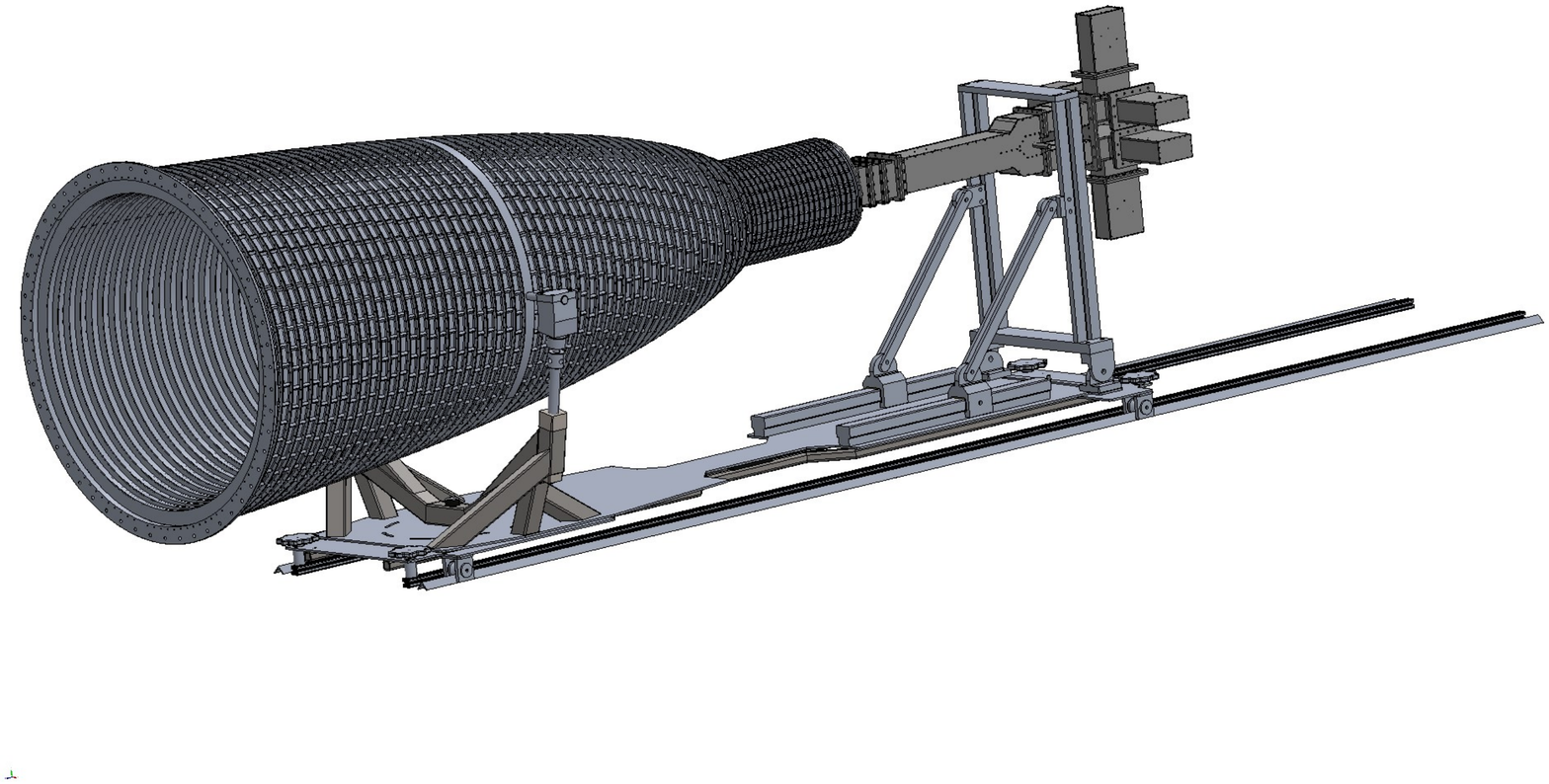}
\caption{Horn support system, designed to move the horn up and down according to the calculated positioning values. In the case of the Double Rectangular and triple rectangular the maximum value is $\pm 30$ cm from a central, reference position. In the case of the Rectangular the value is $\pm 42$ cm. \textcolor{black}{This cell support allows for the positioning  of all horns in their optimal location as described in this paper.} This is further discussed in the companion paper II \citep{2020_BINGO_Instrument}.} 
\label{fig.hexagon}
\end{figure}

In order to solve for these two issues, we investigated three other arrangements outlined in this same figure and discuss the benefits of each configuration in the remaining of this section. The three other configurations outlined in Fig. \ref{fig:FocalPlaneDesign}, called \emph{Rectangular} (top left), \emph{Double Rectangular} (top right) and \emph{Triple Rectangular} (bottom left) have larger cells, allowing the horns to be displaced upwards and downwards within the cell by up to 30 cm. The structure that encapsulates each horn has 2.4 m height and, given their larger size, allows the original horn  positions to be adjusted in the vertical position (referring to the encapsulating cell enclosure centre) by $\pm15$ and $\pm30$ cm, changing the position each horn illuminates the secondary mirror. For the specific case of the Rectangular arrangement we used $\pm21$ cm and $\pm42$ cm for the shifts\footnote{You can access the notebook that created this configuration on this \href{https://github.com/marinscosmo/BINGO_OpticalDesign/blob/main/DrawnModels.ipynb}{link}.}. Fig. \ref{fig.hexagon} shows the mechanism where the horn is attached to the hexagonal case.

These three setups (displacements) allow for the necessary height adjustments to produce a very uniform sky coverage, in a slightly larger declination interval. In the Fig. \ref{fig:FocalPlaneDesign}, there is no displacement, all horns are centered.  With the extra height of the hexagons we note from the figure that the focal plane for all three rectangular configurations is slightly larger than 16 meters required in order for us to obtain a larger coverage than the 15 degrees in declination. A full simulation of the integrated beams over right ascension is presented in the results section and show indeed that some of these configurations are more optimal than others. 

In the \emph{Double Rectangular} configuration 
two rows of detectors are shifted compared to the original first two rows of detectors by a one quarter height of the hexagon height. In other words, the height in the third and fourth rows of the \emph{Double Rectangular} array are shifted upwards from the first two columns by 60 cm. Given that the hexagonal configuration indicates that the difference in horn heights between the first and the second columns is half the hexagonal height, i.e., 120 cm, we therefore have a configuration where each four horns cover the declination by being shifted by 60 cm steps in the $x$ direction compared to one another. We can therefore reach a better than Nyquist configuration in this array by simply shifting the position of the horns once during the survey lifetime. 

Should this be done each year of the survey, we can obtain a map that is over-sampled in the $x$ direction compared to Nyquist sampling, being able to even explore the use of other techniques to extract further resolution from the maps such as the drizzle technique used in the \textit{Hubble Space Telescope}  \citep{2002PASP..114..144F}. The \emph{Triple Rectangular} array 
is composed of more horns and is a configuration that is possible should there be the budget to have more horns in the focal plane. It is similar to the \emph{Double Rectangular} with the caveat that the columns are shifted by one third of the spacing between the hexagonal cells.


In the next sections we make full simulations for the four arrangements shown in Fig. \ref{fig:FocalPlaneDesign} to ensure that the horns are in optimal locations of the focal plane of the telescope and show with full electromagnetic simulations the effects of the focal plane distribution as well as the implications of these choices on the beam shapes and polarisation of the beam.

\section{Optimisation of the focal plane}
\label{sec:optimasation}

Given that obtaining the best solution for all points in the focal plane is a long and time consuming computational task, we outline here a fitting technique where we select points in the focal plane (not in relation to the positions of the four arrangements, but for points that cover all regions of interest, which cover those arrangements) where we fully solve for the best possible locations and angles of the horns in the focal plane and then interpolate for this solution on top of the solved points for each horn position in our focal plane arrangements from the last section. 

For all the four models for the configuration of the horns 
tests were carried out in order to identify their best positioning both in rectangular coordinates ($x, y, z$) and in angular coordinates ($\theta$, $\phi$)\footnote{\textcolor{black}{$\theta$ and $\phi$ describe the pointing of the feed, (and $\psi$ describe  the polarisation direction). They can be described as follows: Let the focal plane system xyz. $\theta$ is angle such that the z-axis rotating around the xy-plane such that the arc described by the z-axis is $\phi$ of the x-axis. So, in the new coordinate system, $\psi$ rotating around the new z-axis by $\psi$.\newline
$\theta$, $\phi$ and $\psi$ angles can be described by Euler angles in the zyz rotating system, that is, first a rotating of $\alpha$ around the z-axis, secondly a rotating of $\beta$ around the y-axis and after $\gamma$ in around the z-axis.\newline
There are association between the angles, given by:
($\theta$, $\phi$, $\psi$) = ($\beta$, $\alpha$, $\alpha + \gamma$)
That we model in a way that $\psi=0$ ($\gamma=-\alpha$). Therefore, $\theta=\beta$ and $\phi=\alpha$.}}. 
Within all but the hexagonal arrangement, the horns can move up to 30 cm upwards and downwards in vertical coordinates on the focal plane. In the arrangements \emph{Double} and \emph{Triple Rectangular} we have no central horn. The \emph{Double Rectangular} was simulated with 28 horns, seven for each column, but there is a possibility to extend this upwards with another 4 horns. In Fig. \ref{fig:FocalPlaneDesign} these additional horns were represented by dashed lines.

We note here that we have optimised the horn locations and angles by looking at the maximum signal intensity which was generated by our simulations at a frequency of 1100\,MHz (roughly at the center of our central frequency of 1120\,MHz. However, the BINGO frequency coverage spams 980\,MHz to 1260\,MHz, so, although the optimisation in this work was performed at this frequency, we also observe the behaviour of the beams in other frequencies between 980\,MHz and 1260\,MHz. Given that we perform this optimisation throughout the focal plane this applied to all arrangements\textcolor{black}{)}.

In order to obtain the ($z$, $\theta$, $\phi$) positions for each horn achieving maximum intensity, we sought the values of these parameters that maximized $10\log_{10}(\|E_{\rm co}\|^2)$ using the TICRA-GRASP software. These parametric values are equivalent to the intensity itself;\footnote{The intensity is obtained by $10\log_{10} (\|E_{\rm co}\|^2+\|E_{\rm cx}\|^2) \approx 10\log_{10}(\|E_{\rm co}\|^2)$ the peak position does not change either. } the same is true for the $Q$ polarisation parameter. The first calibrations were made as follows: given a horn to be analyzed, identified by positions $x$ and $y$, we looked for the value of the parameters $(z,\theta, \phi)$  that provided the maximum amplitude, scanning all the possible values. This has been done for a set of 52 $(x,y)$ positions.

The data obtained by the above method made it possible to fit for a $z$ as well as for a $\theta$ symmetry, and a $\phi$ anti-symmetry with respect to $y$. We thus achieved a total of 96 positions for our parameter fits (all these 96 positions are obtained independently of the arrangements discussed). We searched  analytic fits to represent $(\hat{z}, \hat{\theta}, \hat{\phi})$, for each $(x,y)$ in the region of interest. It was possible to find good fits (regression model with non-linear least squares) for the angular parameters using only a polynomial as a function of $x$ and $y$. We use a function $\hat{z}(x,y)$, $\hat{\theta}(x,y)$, $\hat{\phi} (x,y)$ defined as

\begin{eqnarray}
a &+& bx    + cy   +  dxy + ex^2 + fy^2 + gx^2y + hxy^2\nonumber\\
  &+& ix^3 + jy^3 + kx^4 + ly^4 + \frac{m}{\left(x+10^{-5}\right)} + \frac{n}{\left(y+10^{-5}\right)}\,,
\label{eqn: fitting_focalplane}
\end{eqnarray}
with coefficients from \textit{a} to \textit{n} presented in Table \ref{tab:fitting_parameters_used}. We note that given that we have performed the fits throughout the focal plane, the fits apply to all arrangements.

\begin{figure*}
\centering
\includegraphics[scale=0.18]{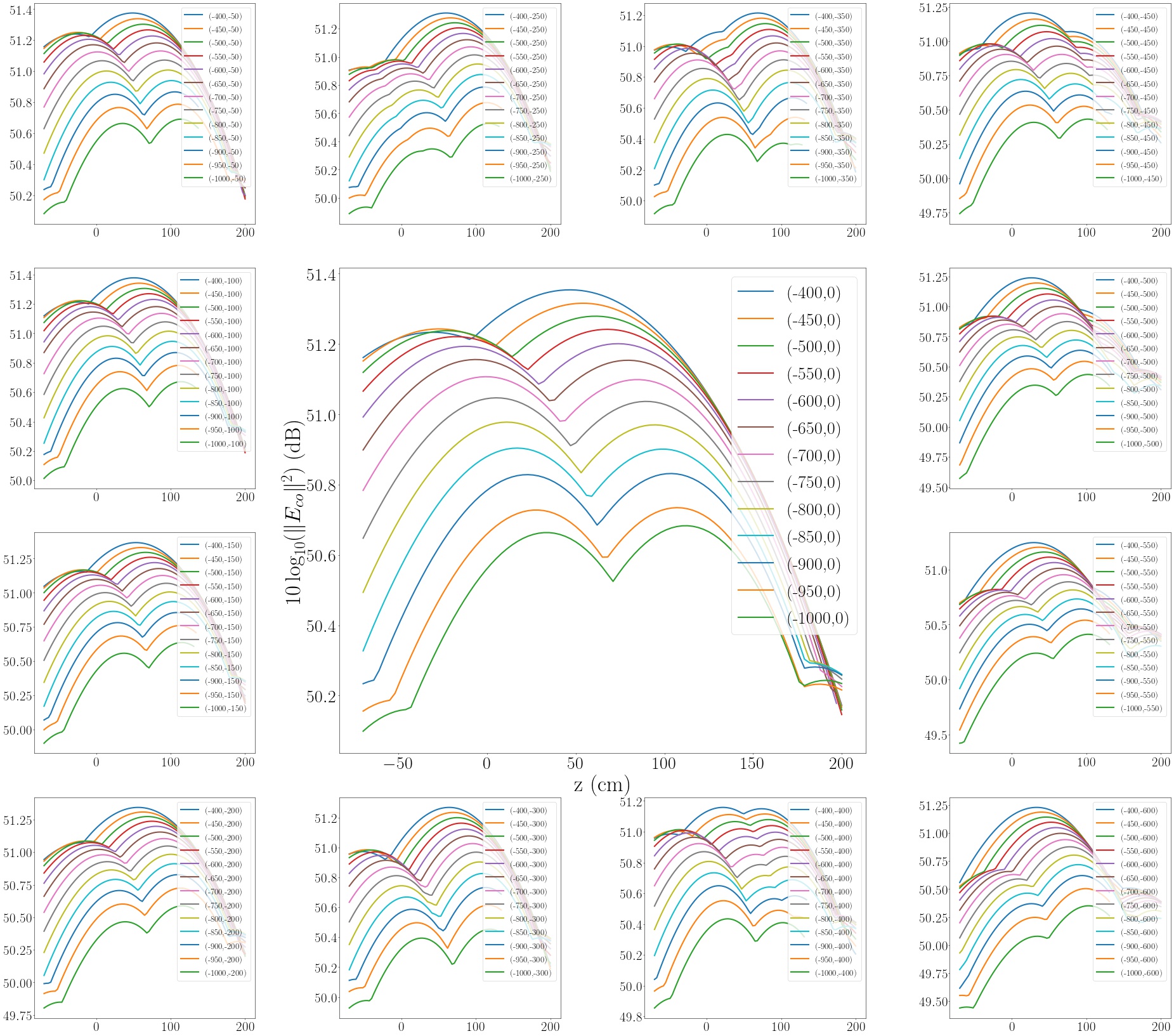}
\caption{Evaluation of $z$ parameter in relation to $x$ coordinate for different $y$ coordinates. The $x$ values are between $-1000$ and $-400$ cm with 50 cm intervals for all plots. The y values are between $-600$ and 0 cm, starting in upper left plot and finishing in lower right plot. These values corresponds to critical region in focal plane for $z$ parameters to obtain the fit.}
\label{fig:evaluation_z}
\end{figure*}

The $z$-fit required a more refined analysis to understand its lack of smoothness, with some positions appearing quite different from others. To analyze this region, with the calibrated values of the angular parameters, we made a complete analysis of how the amplitude varies with $z$. Taking the maximum intensity for $x>-400$ cm values generated a gradual evolution of the parameter, but not for $x<-400$ cm. The amplitude can be seen at different positions in Fig. \ref{fig:evaluation_z} where each graph represents the amplitude with respect to $z$ for a given $y$ and $ x $ spaced  by 50 cm, from $-400$ to $-1000$ cm. A table of the results for this optimisation is given in Appendix \ref{AppendixA}.

\textcolor{black}{We note here that our optimization procedure creates an optical surface where the horns will be placed. This optical surface has a small curvature, as can be seen in figure \ref{fig:grasp_python}. This optimization leads to an optical surface that is continuous, positioning the horns in an optimal way. The horns are not expected to see each other much more than they would if placed side by side on a plane. However, we do not take into account the coupling between the horns when located in the optical surface. This is currently assumed to be small and will investigated in future modelling.}

\begin{figure}
    \centering \includegraphics[scale=0.36]{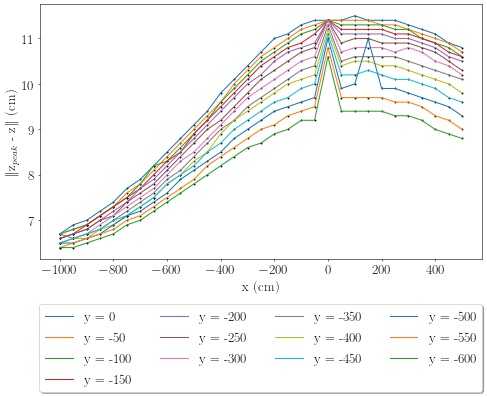}
    \caption{\textcolor{black}{Depth of field needed for appropriate focusing of the horn, considering a defocusing of $0.005 dB_i$. Let $z_{max}$ be the z value for peak of the intensity, $\|z_{max}-z\|$ are the (absolute) values corresponding to $\|I_{max}- I\|\ =\ 0.005$ dBi. Each plot correspond to a specific y value in the focal plane for different x values, that is, representing the value of $\|z_{max}-z\|$ through a vertical path in the focal surface. Smaller (negative) x values correspond to smaller $\|z_{max}-z\|$ values. The second peak in the y=-500 curve (blue), in $x=150$ position is because the evaluation with I(x,y,z)=I(150,-500,z) curve has a division where there would be a peak, around z = -8 cm. Because of that, the new peak has a smoother declination and then greater $\|z_{max}-z\|$ for 0.005 dBi.  That division corresponding to the division z $\sim$ 74 cm for I(x,y,z)=I(250,-500,z) and z $\sim$ 78 cm for I(x,y,z)=I(350,-500,z). This corresponds to our target accuracy with which we need to place the horns in the focal plane.}}
\label{fig:depthoffocus}
\end{figure}

 It is possible to see that for a given $y$, the peak at $x=-400$ cm, is not necessarily the highest peak. This fact is very clear in the graphs for $y=-350$, $-400$ and $-450$ cm. They could be a second or third peaks in the response curve, most likely due to different aberrations being more dominant, and for this reason it was not possible to achieve a monotonous representation of $z$, in the $x$ and $y$ plane, taking only the maximum amplitude values. To have a smooth representation and thus achieve a fit, it was necessary to restrict ourselves to the same peak for different $ x$'s and follow its modification in relation to $ x$, for a given $y$. \textcolor{black}{It is possible to see the depth of focus for all 13 settings above in Fig. \ref{fig:depthoffocus} where we take the intensities at each selected peak and its respective z value called by $I_{max}$ and $z_{max}$, respectively; then, we find $z_{max}-z$ where $I_{max}-I\ =\ 0.005$ dBi and plot it in Fig. \ref{fig:depthoffocus}, where each line represents a fixed y value for variable x positions. We see that all z-values are in the range $6.5 \le y \le 11.5$ cm.}
 
\begin{figure*}[!ht]
    \centering
    \includegraphics[scale=0.29]{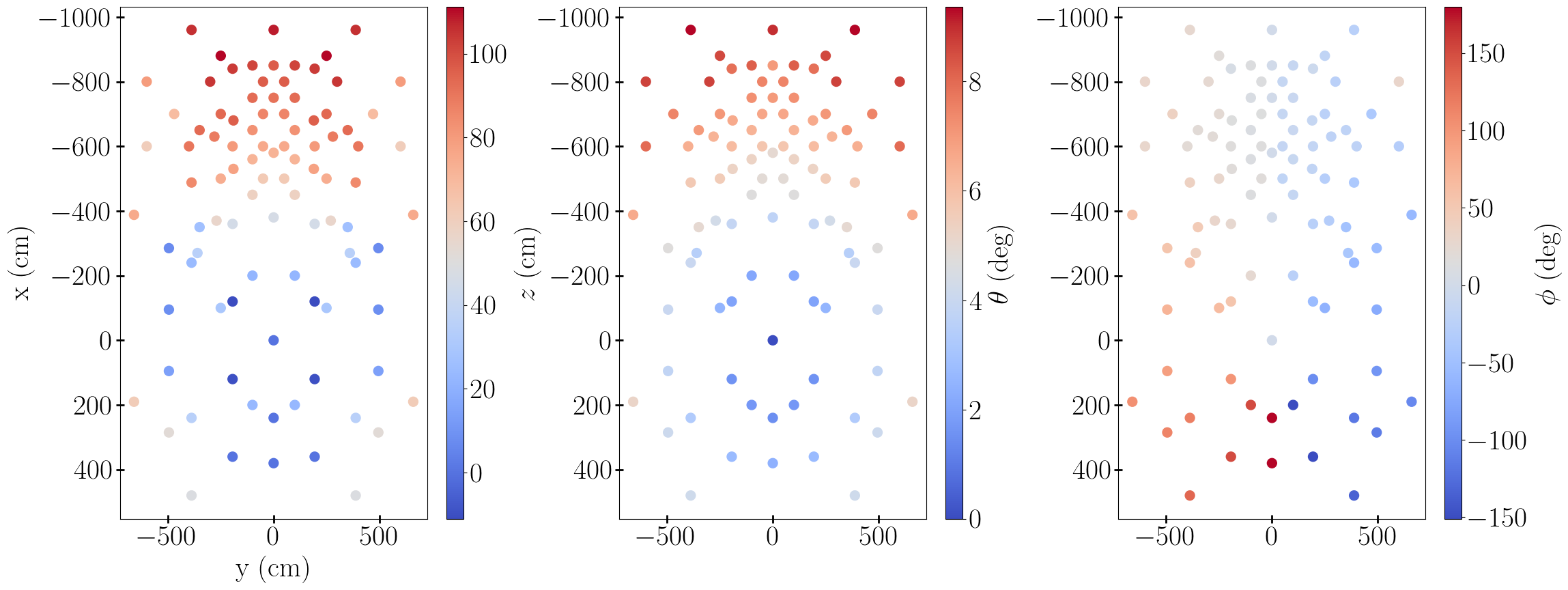}
    \caption{Optimized horns parameters with GRASP only using the same peak in $z$ range in relation to the $x$ coordinate. The vertical axis of the plots are $x$ values in cm and horizontal axis, $y$ in cm.}
    \label{fig:calibrated}
\end{figure*}

 Therefore, we got the $z$ for these peaks in different $x$’s, in the given $y$, and we looked at a smooth function (Eq. (\ref{eqn: z_fit})), that for each $y$; and with the results, the $z$ value that would give us a smooth evolution. With these results, i.e, by taking occasionally a secondary peak to ensure the smoothness of  the function $z$ we have obtained the positions in Table \ref{tab:general_table}, and Fig. \ref{fig:calibrated}. With this fit, we obtain an r.m.s. for the fit of the $z$ position of below 12 cm,
\begin{equation}
\hat{z}_y(x)= a + bx + cx^2 + dx^3 + ex^4 + \frac{f}{\left(x+10^{-5}\right)}+ \frac{g}{\left(x+10^{-5}\right)}\,.
\label{eqn: z_fit}\
\end{equation} 

\begin{figure*}
    \includegraphics[width=18cm,height=5cm]{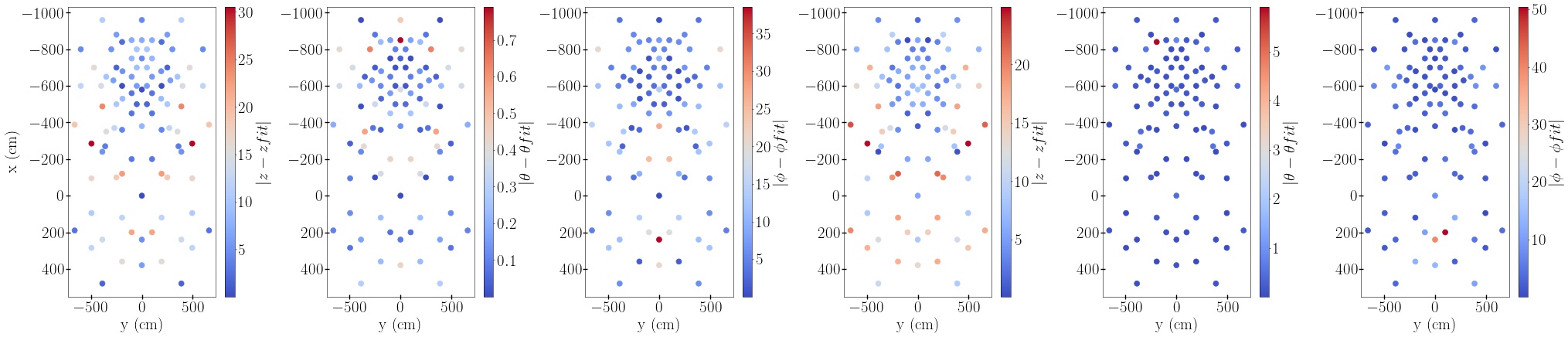}
    \caption{Left: Absolute difference between calibrated parameter values and fitting parameters. The first plot is the difference for $z$ parameter, the centre plot is the difference for $\theta$ parameter and the last plot for $\phi$. The fit used was that of Eq. (\ref{eqn: fitting_focalplane}) with values in Table \ref{tab:fitting_parameters_used}. The vertical axis of the plots are $x$ values in cm and horizontal axis, $y$ in cm. Right: the same analysis, but the fit used was computed by the DNN. 
    }
\label{fig:diff_horns}
\end{figure*}

The $z$-r.m.s. is 11.63 cm; for $\theta$ we found 0.25 degrees and for $\phi= 10.27$ degrees. In Fig. \ref{fig:diff_horns} we can see the absolute difference between the calibrated values and those reconstructed in each position.

\begin{figure}
    \includegraphics[scale=0.27]{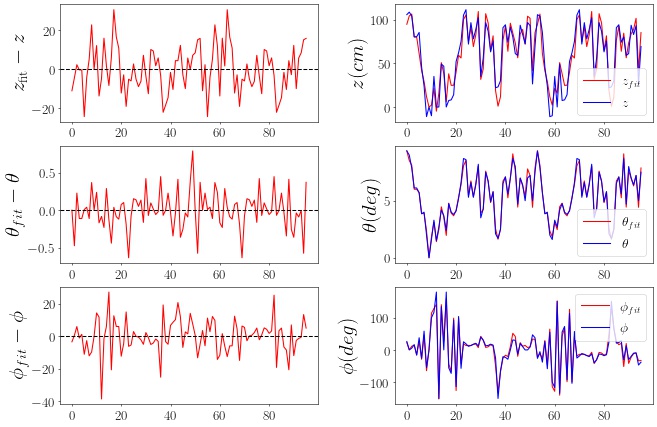}
    \caption{Left column: difference between optimized and fit ($z, \theta, \phi$) values for 96 positions on the focal plane. Right column: comparison between both results.}
    \label{fig:RGxDNN1}
\end{figure}

\begin{figure}
    \includegraphics[scale=0.27]{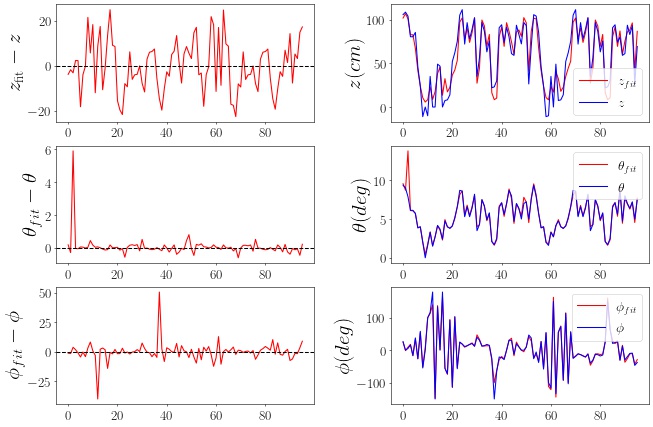}
    \caption{Same as in fig. \ref{fig:RGxDNN1}, but computed with a DNN fit. }
    \label{fig:RGxDNN2}
\end{figure}


We also build a dense neural network (DNN) to reconstruct the parameters $z$, $\theta$ and $\phi$. Thus we build $3$ DNN, using the data obtained by the GRASP simulations.  For the $3$ parameters, we choose a DNN with $10$ neurons. First we divide our dataset in train and test sets where this division is random to avoid overfitting. The division was $70\%$ train and $30\%$ test. We also did the standardization by removing the mean and scaling to unit variance. 

Because of the size of the dataset, we made the choice to run $300$ DNN realizations, we have then averaged the outputs for all of our realizations. For each realization we did required that the neural network took $20,000$ minimisation steps to fit our parameters.


The fitting of the $z$ parameter was chosen using the $x$ and $y$ as inputs. Therefore given the position in the $x$ and $y$ we can predict the $z$, so within the training we achieve in the predicted values given by the DNN. Performing this fitting process with several different DNN for the parameters $\theta$ and $\phi$, yields slightly different results. 

After the training is performed, we compute the r.m.s. of our realization over all 300 DNNs, which is provided in Table \ref{tab:MAE_fit}\footnote{That calculation can be seen on this \href{https://github.com/marinscosmo/BINGO_OpticalDesign/blob/main/CoordinateCalibration_and_DesignModels.ipynb}{notebook link}.}. 

At the end, we showed the results obtained by DNN and the non-linear least squares model, which are summarised in Figs \ref{fig:calibrated}, \ref{fig:diff_horns}, \ref{fig:RGxDNN1} and \ref{fig:RGxDNN2}\footnote{The results applied to all the arrangements in this article studied can be accessed on this \href{https://github.com/marinscosmo/BINGO_OpticalDesign/blob/main/CoordinateCalibration_and_DesignModels.ipynb}{notebook link}.}. The above two fits define the best locations for the horns to be placed in the focal plane, which is indeed not a plane but a complex two dimensional curved surface which we have outlined in this section.

For the next calculations of this paper, we used non-linear least squares fit.

\begin{table*}
\centering
\caption{\textcolor{black}{Fitting parameters of the Eq. (\ref{eqn: fitting_focalplane}) used to compute the coordinates $z$, $\theta$ and $\phi$ via a non-linear least squares fit. The fits are compatible with the residuals shown in Fig. \ref{fig:RGxDNN1}}}
\label{tab:fitting_parameters}
\begin{tabular}{ |P{0.1cm}||P{3.2cm}|P{4cm}|P{3.55cm}| }
\hline
 \multicolumn{4}{|c|}{Fitting parameters}\\
 \hline
 & z (cm) & $\theta$ (deg) & $\phi$ (deg) \\
\hline
a & \phantom{$-$} 1.1 $\pm$ 4.2 & 1.6079 $\pm$ 0.0904 & \phantom{$-$} 89.8 $\pm$  3.69 \\
b & $-$ 0.048 $\pm$ 0.013 & $-$ 2.422e-03 $\pm$ 2.80e-04 & \phantom{$-$} 2.183e-01 $\pm$ 1.14e-02 \\
c & \phantom{$-$} 0.00e-03 $\pm$ 9.7e-03 & \phantom{$-$} 0.00e-04 $\pm$ 2.11e-04 & \phantom{$-$} 0.00e-03 $\pm$ 8.42e-03 \\
d & \phantom{$-$} 0.00e-05 $\pm$ 1.7e-05 & \phantom{$-$} 0.00e-07 $\pm$ 3.79e-07 & \phantom{$-$} 0.00e-05 $\pm$ 1.54e-05 \\
e & \phantom{$-$} 2.1e-04 $\pm$ 2.2e-05 & \phantom{$-$} 9.747e-06 $\pm$ 4.82e-07 & \phantom{$-$} 4.52e-05 $\pm$ 1.97e-05 \\
f & \phantom{$-$} 7.6e-04 $\pm$ 4.0e-04 & \phantom{$-$} 9.574e-06 $\pm$ 8.73e-07 & $-$ 2.32e-05 $\pm$ 3.57e-05 \\
g & \phantom{$-$} 0.00e-08 $\pm$ 2.8e-08 & \phantom{$-$} 0.00e-10 $\pm$ 6.14e-10 & \phantom{$-$} 0.00e-08 $\pm$ 2.40e-08 \\
h & \phantom{$-$} 1.97e-07 $\pm$ 3.0e-08 & \phantom{$-$} 4.8107e-09 $\pm$ 6.53e-10 & $-$ 1.941e-07 $\pm$ 2.67e-08 \\
i & \phantom{$-$} 3.9e-08 $\pm$ 6.7e-08 & \phantom{$-$} 2.07e-09 $\pm$ 1.45e-09 & $-$ 3.071e-07 $\pm$ 5.94e-08 \\
j & \phantom{$-$} 0.00e-08 $\pm$ 3.0e-08 & \phantom{$-$} 0.00e-10 $\pm$ 6.63e-10 & \phantom{$-$} 0.00e-08 $\pm$ 2.67e-08 \\
k & $-$ 1.08e-10 $\pm$ 6.1e-11 & $-$ 2.79e-12 $\pm$ 1.31e-12 & $-$ 2.293e-10 $\pm$ 5.36e-11 \\
l & \phantom{$-$} 5.4e-11 $\pm$ 9.4e-11 & $-$ 4.10e-12 $\pm$ 2.02e-12 & $-$ 2.56e-11 $\pm$ 8.27e-11 \\
m & \phantom{$-$} 5e-05 $\pm$ 1.4e-04 & $-$ 1.416e-05 $\pm$ 2.99e-06 & $-$ 9.14e-04 $\pm$ 1.22e-04 \\
n & $-$ 6.3e-05 $\pm$ 5.5e-05 & $-$ 1.92e-06 $\pm$ 1.19e-06 & \phantom{$-$} 1.67e-05 $\pm$ 4.85e-05 \\
\hline
\end{tabular}
\label{tab:fitting_parameters_used}
\end{table*}

\begin{table}
\centering
\caption{Root Mean Square of the coordinates computed in table \ref{tab:fitting_parameters_used} for the non-linear least squares fit and dense neural network (DNN) fits.}
\begin{tabular}{ |P{1.5cm}||P{2.cm}|P{1.5cm}| }
 \hline
Parameter & non-linear LS & DNN\\ 
\hline
z (cm)        & 11.63 & 10.21\\
$\theta$ (deg) & \phantom{0}0.25  & \phantom{0}0.68\\ 
$\phi$ (deg)  & 10.27 & \phantom{0}7.89\\ 
\hline
\end{tabular}
\label{tab:MAE_fit}
\end{table}

\section{Results}\label{sec:results}

\begin{figure*}
    \centering
    \begin{subfigure}[b]{.45\linewidth}
        \includegraphics[width=\linewidth]{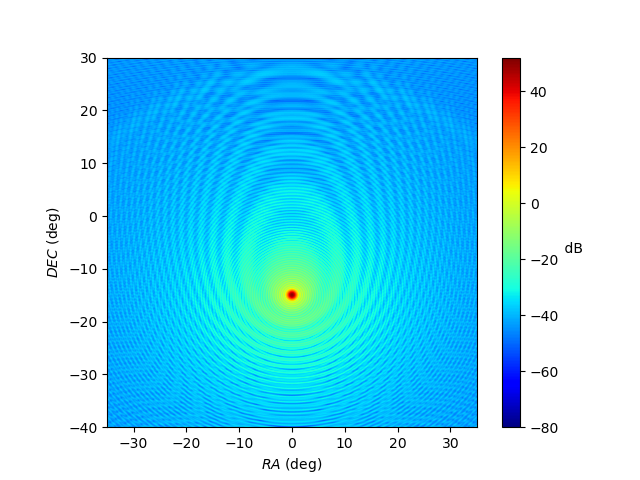}
        \caption{Monochromatic}
        \label{fig.sidelobe_mono1110}
    \end{subfigure}
    \begin{subfigure}[b]{.45\linewidth}
        \includegraphics[width=\linewidth]{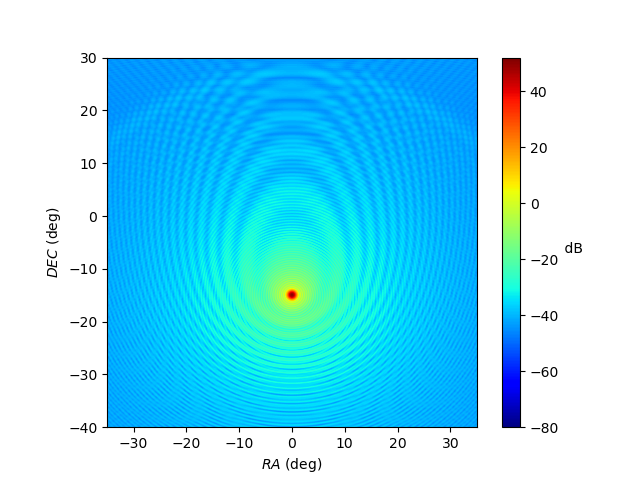}
        \caption{10\,MHz width}
        \label{fig.sidelobe_10width}
    \end{subfigure}
    \begin{subfigure}[b]{.45\linewidth}
        \includegraphics[width=\linewidth]{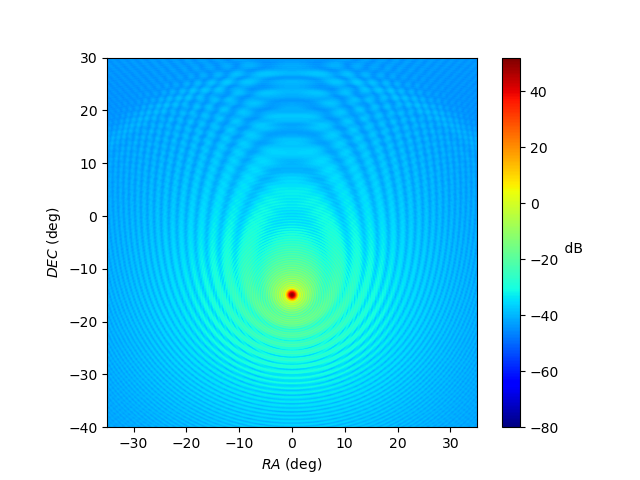}
        \caption{20\,MHz width}
        \label{fig.sidelobe_20width}
    \end{subfigure}
    \begin{subfigure}[b]{.45\linewidth}
        \includegraphics[width=\linewidth]{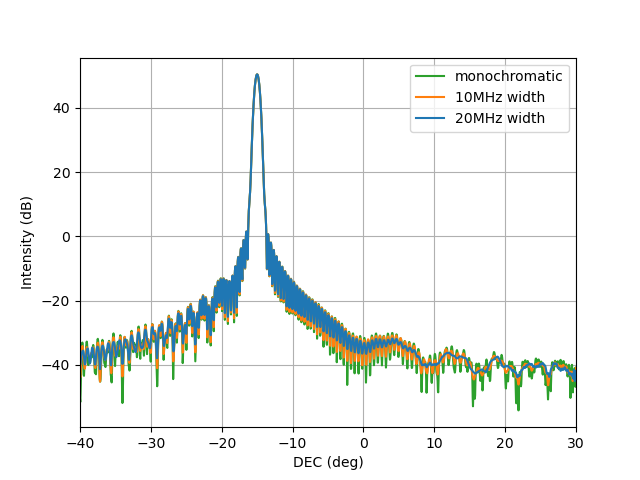}
        \caption{1-D}
        \label{fig.sidelobe_1d}
    \end{subfigure}
    \caption{The side lobes of the BINGO central beam present a Moirré effect due the two-mirror configuration of the instrument (Figure~\ref{fig.sidelobe_mono1110}). We found that this effect is attenuated when integrating the beam. Figures \ref{fig.sidelobe_10width} and  \ref{fig.sidelobe_20width} show integrated beams over widths of 10\,MHz and 20\,MHz respectively, computed in the range $1105 -1115$\,MHz and $1100  - 1120$\,MHz, with displacements of 1\,MHz and normalised by the number of steps. The attenuation effect is  clearer in Fig. \ref{fig.sidelobe_1d}, portraying a 1-dimensional cut of the results from figures~\ref{fig.sidelobe_mono1110}),  \ref{fig.sidelobe_10width} and  \ref{fig.sidelobe_20width}. } 
    \label{fig.sidelobe}
\end{figure*}

\subsection{The beam analysis}

In this section we describe the computation of the beam profiles for each arrangement we have previously considered. We use the GRASP software for any horn of 
any given arrangement where each one is defined by their standard rectangular and angular coordinates, both of which have been computed and interpolated using the previously described Least Squares Fitting of the optical surface. GRASP propagates a monochromatic, Gaussian and linearly polarised beam through the optical system, and provides the response as a 2-dimensional complex electrical field 
in $u-v$ coordinates over the sphere. We choose, however, to present the results in celestial coordinates (RA, DEC) chosen at a given specific time of the survey (specified as any time when RA of zero is located directly south), given that BINGO is a transit telescope. 
For this purpose we perform a linear interpolation of the data according to the sequence of transformation defined below. First of all, we transform (u,v) to standard Cartesian coordinates ($x$, $y$, $z$) on the unit sphere
\begin{align}
    x &= u\,, \\
    y & = v\,, \\
    z &= \sqrt{1 - u^2 -v^2}\,.
\end{align}

The $u-v$ coordinates can be understood as a plane tangent to the sphere on the center of the beam, which is at $x=y=0$ and $z=1$ in this system. In order to transform to Celestial coordinates, however, it is helpful to define a new Cartesian system such that the $z$-axis points to the Celestial North pole. We choose to rotate the center of the focal plane to be located at declination $-15^{\circ}$ of the Celestial Equatorial plane, therefore we are required to perform a rotation of $105^{\circ}$ (i.e. $90^{\circ}$+$15^{\circ}$) from the celestial pole, which is where GRASP was set up to produce our beam:
\begin{align}
    \begin{pmatrix} x' \\ y' \\ z' \end{pmatrix} &= 
    \begin{pmatrix} \cos 105^\circ & 0 &  \sin 105^\circ \\
    0 & 1 & 0 \\
    -\sin 105^\circ & 0 & \cos 105^\circ
    \end{pmatrix}
    \begin{pmatrix} x \\ y \\ z \end{pmatrix}
\end{align}

\noindent and the definitions of the Celestial coordinates follow straightforwardly
\begin{align}
   \text{ DEC} &= \text{arcsin}(z')\,,\\
    \text{RA} &= \text{arctan2}(y',x') = 2\,\text{arctan}\left( \frac{y'}{\sqrt(x'^2 + y'^2) + x'} \right)\,.
\end{align}
Fig. \ref{fig.sidelobe} shows the response due to the BINGO central horn only. The quantity being plotted is the intensity $I_{\rm dB} = 10 \log_{10} (I)$, where the intensity $I$ (in watt) is defined by the sum of squares of the two orthogonal components of the electrical field
\begin{align}
    I = |E_{\rm co}|^2 + |E_{\rm cx}|^2\,.
\end{align}

\noindent where $E_{\rm co}$ is the electric field measured in linear polarisation and $E_{\rm cx}$ is the electric field measured in cross polarisation. For the single horn analysis, the number of points $N_p$ for writing the fields outputs of the GRASP software was chosen to be $1614 \times 1614 $ in a ($u,v$) range from $-0.7$ to $0.7$. 
We further compute the average of the intensities from each linearly polarized beam.

\begin{figure}[!ht]
    \centering
    \includegraphics[scale = 0.39]{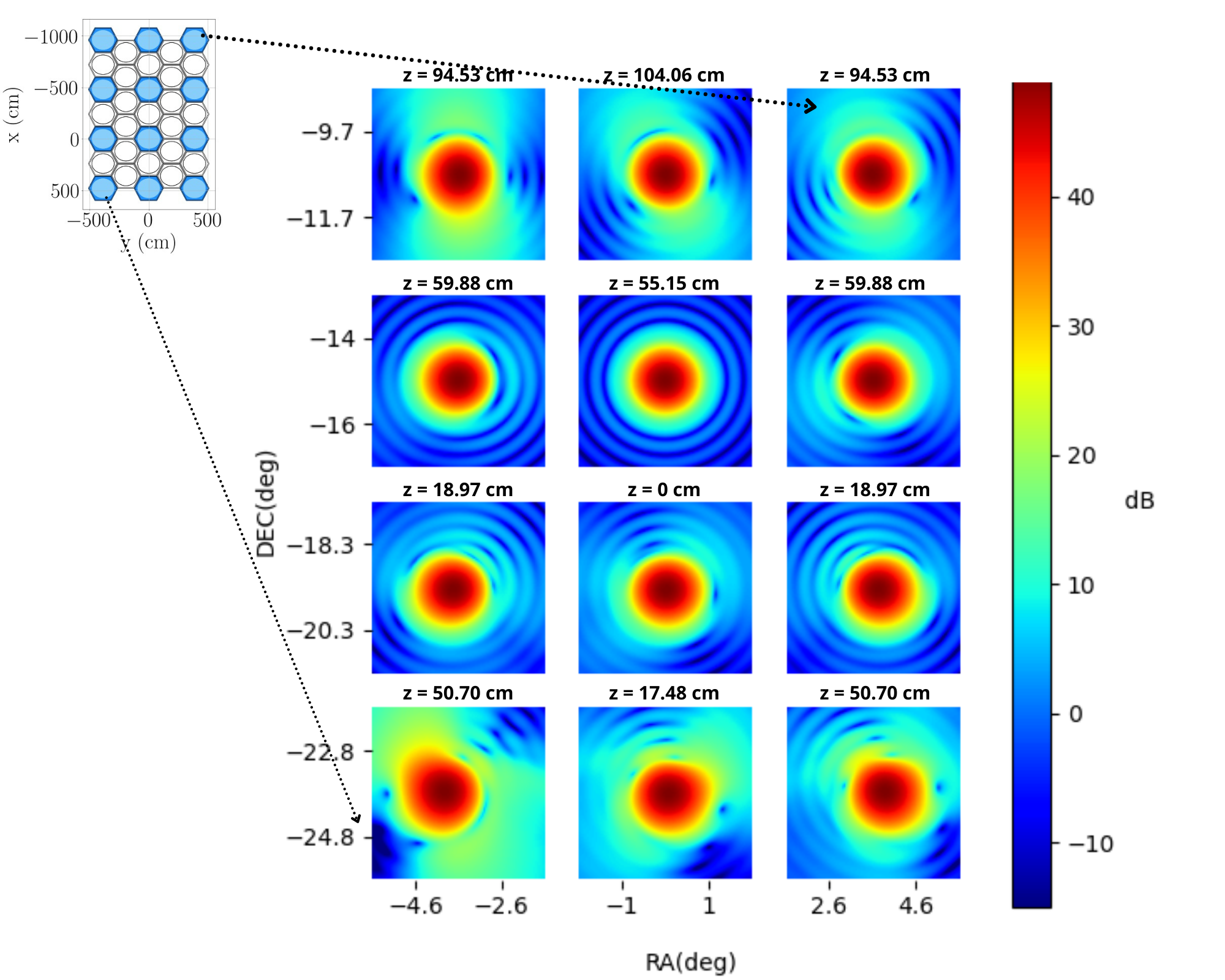}
    \caption{Optical aberrations for the final beams of the horns located at different places within the focal plane (\textcolor{black}{corresponding locations x and y chosen according to upper left figure, and the respective z values over each horn, in the focal plane localization}). The focus of the telescope is placed at $\delta = -15^{\circ}$ and has a beam that is devoid of aberrations while the aberrations are increasingly large, but less than $\sim 10 dB$, as we move away from the center of the focal plane.}
    \label{fig.beams12}
\end{figure}

Figure~\ref{fig.beams12} shows the optical aberrations for the beams of a hypothetical focal plane, with the central horn located at the focus of the telescope, which points at $\delta = -15^{\circ}$. Due to the horn arrangement, the boresight is always at a declination $\delta < \delta_{center}$. Nevertheless, we can see that the beam shape is relatively well behaved across this hypothetical focal plane, and that the aberrations remain at a level below 30\,dB, at $ \textrm{(RA, DEC)} \lesssim |0.67|^{\circ}$ from the center of the main beam. This gives us confidence that the optical aberrations within the field of view chosen by this work can be well modelled when the final survey is produced. Table \ref{tab.fwhm} shows the FWHM values for the beams presented in Fig.\ref{fig.beams12}. Values are listed in the same order as beams in the figure. \textcolor{black}{We also calculated the ellipticity, defined as $e = \sqrt{1-((FWHM_x)/(FWHM_y))^2}$ for a set of 5 horns located at several locations in the focal plane at positions $(x, y)$: ((0,0),(450,304),(510, -304),(-930,-304),(-990,304)). In all sets, the calculated ellipticity values $\epsilon$ are smaller than  0.1, meeting the requirements from \citet{2012arXiv1209.1041B}.}

\begin{table}[ht]
\caption{FWHM (deg) of beams computed in Fig. \ref{fig.beams12}}
\label{tab.fwhm}
\centering
\begin{tabular}{ ccc } 
 \hline
0.633 & 0.611 & 0.615 \\
0.601 & 0.592 & 0.607 \\
0.620 & 0.632 & 0.619 \\
0.711 & 0.671 & 0.663 \\
 \hline
\end{tabular}
\end{table}

Furthermore, for the sake of comparing the different configurations, we compute all beams of each configuration for frequencies of 980\,MHz, 1100\,MHz and 1260\,MHz setting accurately the attenuation angle according to the method previously explained in Section \ref{sec:taper}. Also, we average the response of each linearly polarised beam. We recall that, although we present the results in $(RA,DEC)$ coordinates, the originally simulations are made in $(u,v)$ coordinates. We choosen range in $(u,v)$ coordinates for the simulations is from $-0.2$ to $0.2$ for the \emph{Hexagonal} and \emph{Rectangular} arrangement, with $461 \times 461$ points, and from $-0.4$ to $0.4$ for the \emph{Double-Rectangular} and \emph{Triple-Rectangular} arrangement with $922 \times 922$ points. Then, although the simulated area is not the same we keep a similar resolution of $1152.5^2$ points per unit of area, which has been shown to be good enough for the interpolation procedure.
Moreover, we define a total intensity for the arrangement, normalized by the intensity peak of the central horn, according to 
\begin{equation}
p^{\nu}_{\rm arr} (\text{RA},  \text{DEC}) = \frac{\sum^{\rm horn} I^{\nu}_{\rm horn} (\text{RA},  \text{DEC})    }{I^{\text{1100 MHz}}(0, -15^\circ)}\label{intensity_normalized_2d}
\end{equation}
and its one-dimensional version as well 
\begin{equation}
P^{\nu}_{\rm arr} (\text{ DEC}) = f \times \frac{ \int d\text{RA} \; \sum^{\rm horn} I^{\nu}_{\rm horn} (\text{RA}, \text{DEC})    }{ \int d\text{RA} \; I^{\text{1100 MHz}}(\text{RA}, -15^\circ)}\,,
\label{intensity_normalized_1d}
\end{equation}
where the RA dependence is removed by integration.

\begin{figure}
    \centering
        \includegraphics[width=7.5cm]{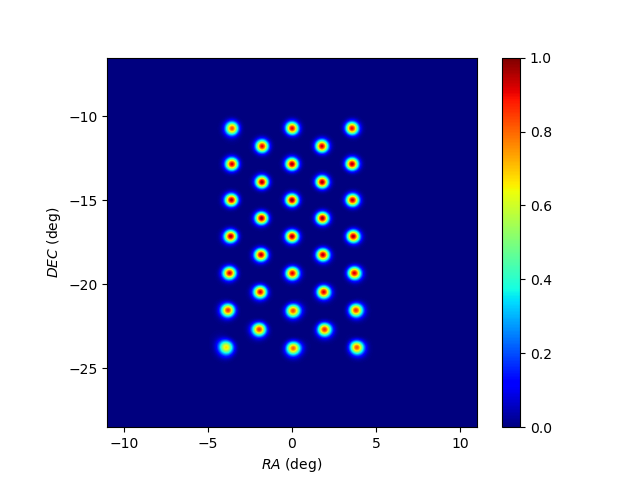}
        \includegraphics[width=7.5cm]{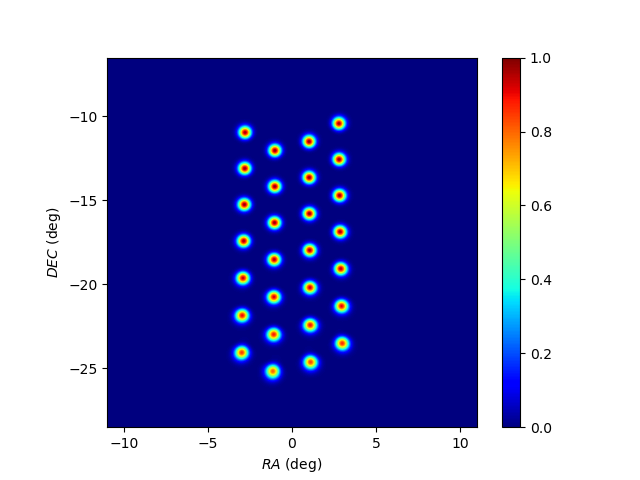}
        \includegraphics[width=7.5cm]{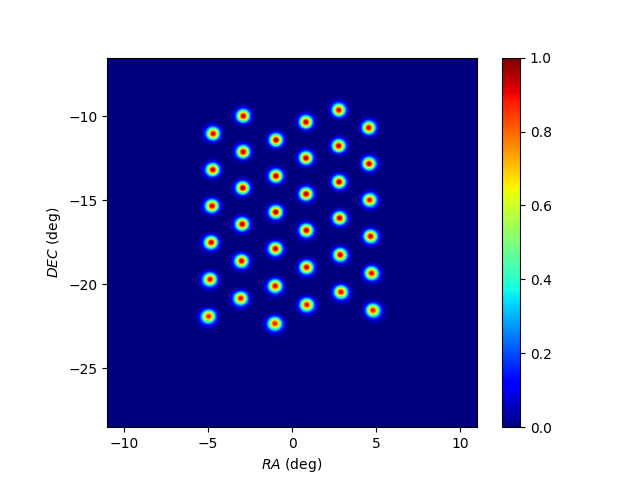}
        \includegraphics[width=7.5cm]{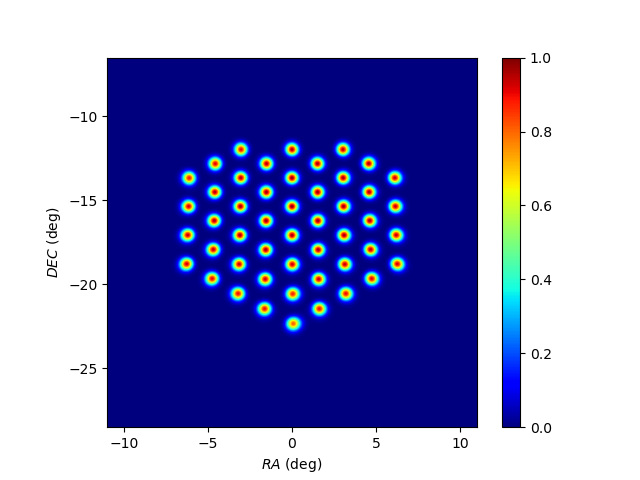}
    
    \caption{Beam responses (intensities, from top to bottom) for:   
    a) \emph{Rectangular}, 
    b) \emph{Double-Rectangular}, 
    c) \emph{Triple-Rectangular} 
    and 
    d) \emph{Hexagonal}  
    normalised by the intensity of the central horn (see Eq. (\ref{intensity_normalized_2d})).}
    \label{fig.arrangements_I}
\end{figure}

\begin{figure}
    \includegraphics[width=9.0cm]{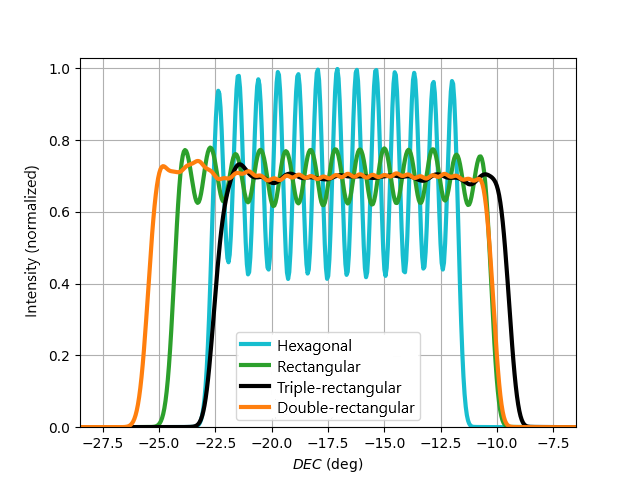}
    \caption{Response function of the beams, integrated in RA, for selected columns (corresponding to a declination range) and different horn arrangements. Results are normalized to the response of the central horn, according to Eq. \ref{intensity_normalized_1d}. For the \emph{Triple-Rectangular} and \emph{Double-Rectangular} arrangements, the beam average also took into account the five displacements of $0, \pm 15$\,cm, $\pm 30$\,cm. For the \emph{Rectangular} arrangement, the five displacements were $0, \pm 21$\,cm, $\pm 42$\,cm. This averaging produces an even smoother profile for the  arrangements. We note that it is possible to change the orientation of the above figures on the sky by changing the north-south orientation of the telescope but maintaining the same boresight. The current configuration was chosen in order have a simpler structure for the support of the mirrors and horns.}
    \label{fig.beam1d}
\end{figure}

The results of applying Eq. (\ref{intensity_normalized_2d}) to the \emph{Double-Rectangular}, \emph{Triple Rectangular}, \emph{Rectangular} and Hexagonal arrangements are shown in Fig. \ref{fig.arrangements_I}, while the results of applying Eq. (\ref{intensity_normalized_1d}) are depicted in Fig. \ref{fig.beam1d}. 
Unless stated otherwise, all results were computed at $f=1100$\,MHz. Although the sum in Eq. (\ref{intensity_normalized_2d}) is over \textit{all} horns, in Eq. (\ref{intensity_normalized_1d}), for the \emph{Rectangular} and Hexagonal arrangement, the selection covers the central column of horns and also one adjacent. For the Double and \emph{Triple Rectangular}, however, we select all the columns, since its particular geometry does not create overlaps. For this reason,  in Eq. (\ref{intensity_normalized_1d}) we set $f=1/2$ for the \emph{Double-Rectangular}, $f=1/3$ for the \emph{Triple Rectangular}, and $f=1$ for the \emph{Rectangular} and \emph{Hexagonal}, since for the Double (and Triple)-Rectangular, the beams are summed up over twice (and triple) the numbers of columns. This has been chosen so that the contents of Fig. \ref{fig.beam1d} reflect the integrated beam while the sky swipes across the focal plane. 

\textcolor{black}{We stress here that we have interpreted the results of the double rectangular distribution as the best results given that the smoothness of the coverage would indicate that the noise properties of the final map are as close to homogeneous as possible. If maps were not Nyquist sampled and the resulting projected beam along the declination direction was not smooth, we would obtain a map which has noise properties which have a pattern in the sky. This is clearly undesirable when it comes to perform a power spectrum analysis of the residuals as we would have to model this in-homogeneity perfectly in order to obtain results about the power spectrum of 21cm radiation. Therefore the smoother versions of such an arrangement are preferred than the non-smooth versions. An analysis of the impact of this on the power spectrum of the final data-sets is performed in our sister paper \citep{2020_sky_simulation}.}

Additionally, the results presented in the Fig. \ref{fig.beam1d} were averaged over the five displacements of $0, \pm 15$\,cm, $\pm 30$\,cm for the \emph{Double-Rectangular} and \emph{Triple-Rectangular} and $0, \pm 21$\,cm, $\pm 42$\,cm for the \emph{Rectangular arrangement}. This is so that the final coverage of the focal plane is optimal with respect to the Nyquist theorem as outlined in Section \ref{sec:focal_plane_arrang}. This averaging produces a smoother intensity profile for those three arrangements compared to the \emph{Hexagonal arrangement} which is not averaged. Also, the \emph{Double} and \emph{Triple-Rectangular} presents even smoother profiles compared to the \emph{Rectangular} arrangement, which can be explained by the particular geometry of the arrangements. We conclude that this smoothness will most likely lead to a sky coverage that avoids gaps in the coverage, which is confirmed in our sister paper \citep{2020_sky_simulation}. 
We conclude here that the configuration that best achieves smoothness with the minimum number of horns is the \emph{Double Rectangular}, and is the configuration that we adopt as the standard configuration for the project.

\subsubsection{The beam polarisation}

We investigate the polarisation properties in this section by computing the four Stokes parameters for each beam and summing them up for each arrangement. We apply the following definition
\begin{align}
   I &= |E_{\rm co}|^2 + |E_{\rm cx}|^2 = |E_{a}|^2 + |E_{b}|^2 = |E_{\rm rhc}|^2 + |E_{\rm lhc}|^2 \,,\\
    Q &= |E_{\rm co}|^2 - |E_{\rm cx}|^2 \,,\\
    U &= |E_{a}|^2 - |E_{b}|^2 \,,\\
    V &= |E_{\rm rhc}|^2 - |E_{\rm lhc}|^2\,, 
\end{align}

\noindent where the subscripts refer to three different basis of the space of Jones vectors. Being $(E_{\rm co}, E_{\rm cx})$ the standard Cartesian basis as defined above, $E_a$ and $E_b$ are defined by a rotation of the Cartesian basis used to define $(E_{\rm co}, E_{\rm cx})$, rotated by $45^\circ$ and, $E_{\rm rhc}$ and $E_{\rm lhc}$ a left hand and right hand basis, defined as follows
\begin{align}
        e_{a/b} &= \frac{1}{\sqrt{2}} \left( \pm e_{\rm co} + e_{\rm cx} \right)\,, \\
        e_{\rm lhc/rhc} &= \frac{1}{\sqrt{2}} \left( e_{\rm co} \pm i e_{\rm cx} \right)\,. 
\end{align}
    
The polarised responses of the beam has a certain is non zero away from the focus of the telescope, presenting either positive or negative $Q$ values around 20dB below the gain  of the unpolarised $I$ beam, when averaged over RA. The levels of $U$ polarization of the beam is a further 5-10 orders of magnitude below this depending on the position on the focal plane. We aim to simulate the polarisation gained through the optical system by an originally unpolarised beam, and assuming the input signal is inpolarised, we average the response of both orthogonal polarised beams. While $Q$ and $U$ are highly dependent on these initial states, the $I$ and $V$ values are nearly the same for the two linearly polarised states. Despite this, this method enable us to subtract any effect due the initial state of polarisation. 

\begin{figure*}[t]
    \centering
    \includegraphics[scale = 1.]{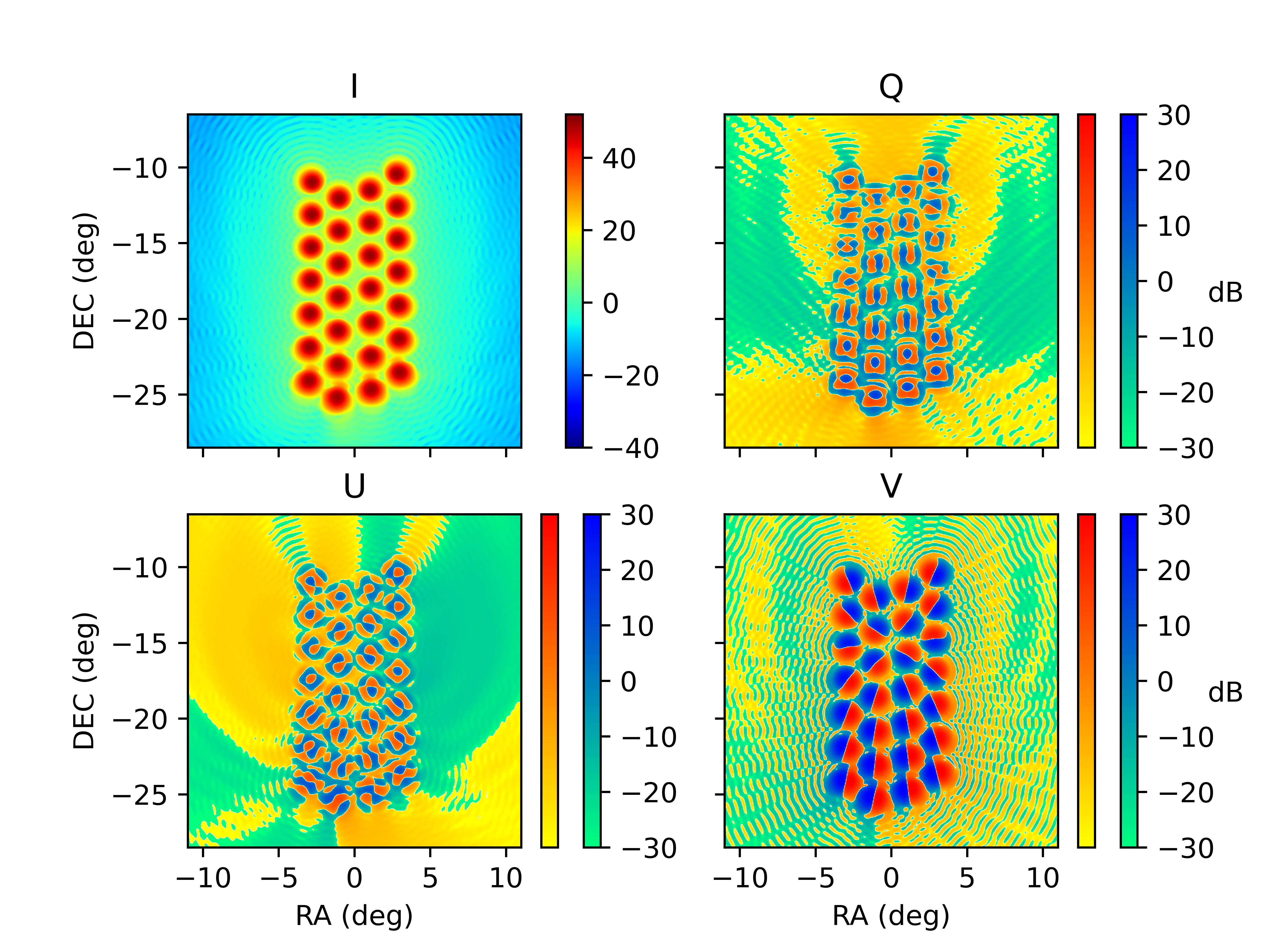}

\caption{ The Stokes parameters $I, Q, U, V$ summed up for the beams of the \emph{Double-Rectangular} arrangement, and transformed to dB units, i.e., $  X_{dB}= 10\log_{10}(\pm X)$. The two color bars indicate the positive (red) and negative (blue) values for the $Q, U$ and $V$ parameters. Each beam was previously averaged between the responses of each linearly polarised state.  }
    \label{fig.stokes_double}
\end{figure*}

\begin{figure}[h]
    \centering
    \includegraphics[width=9cm]{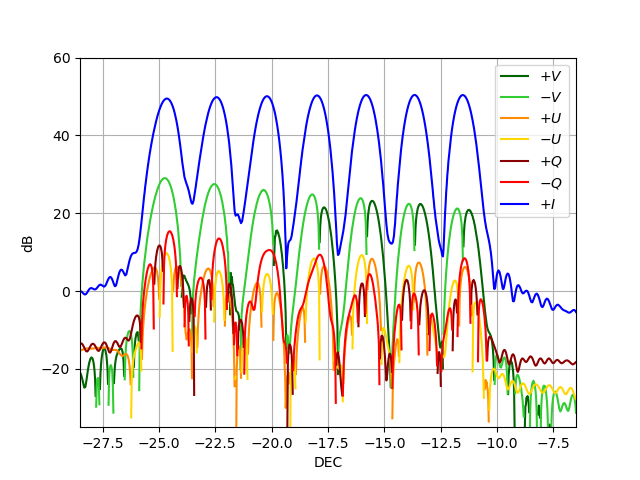}
    \caption{An one-dimensional cut of the Stokes parameters presented in Fig. \ref{fig.stokes_double} for the \emph{Double-Rectangular} arrangement. The cut is over RA $= $RA$_{\rm peak}$, where RA$_{\rm peak}$ is the coordinate of the peak in intensity.  We see that $V$ is at least 25\,dB lower than $I$ on the peaks, while $Q $ and $U$ are even lower.  }
    \label{fig.stokes_param_dec}
\end{figure}

The results of the Stokes parameters for the \emph{Double-Rectangular} arrangement are shown in Fig. \ref{fig.stokes_double}, and also a 1-dimensional cut in Fig. \ref{fig.stokes_param_dec}. Each parameter is summed up for all the beams of the configuration, computed in the frequency of 1100\,MHz. Of course $Q$, $U$, $V$ are allowed to be either positive or negative, and the variation in their sign reveals a balance between the field directions. It is also instructive compare the magnitude of the parameters. In Fig. \ref{fig.stokes_param_dec} we see that $V$ is at least 25\,dB lower than $I$ in the peaks and the difference is even higher for the $Q$ and $U$ parameters. \textcolor{black}{We note here that the results in Fig.\ref{fig.stokes_double} represent the response of unpolarized point sources located where each of the horns are pointing in the sky, therefore represent the optical contribution of the leakage from the arrangement, including from side-lobes, but it neglects any other potential leakage that might be included in the entire rest of the instrument, including the electronics.}

\subsubsection{The beam window function.}

In this section we compute the power spectra for the beams to investigate the effect it will have in modifying the final power spectrum of the 21-cm radiation that we are studying. It is important to estimate this as any strong features in the beam window function could be misinterpreted by a measured feature in the 21cm power spectrum.

\begin{figure}[h]
    \centering
    \includegraphics[width=9cm]{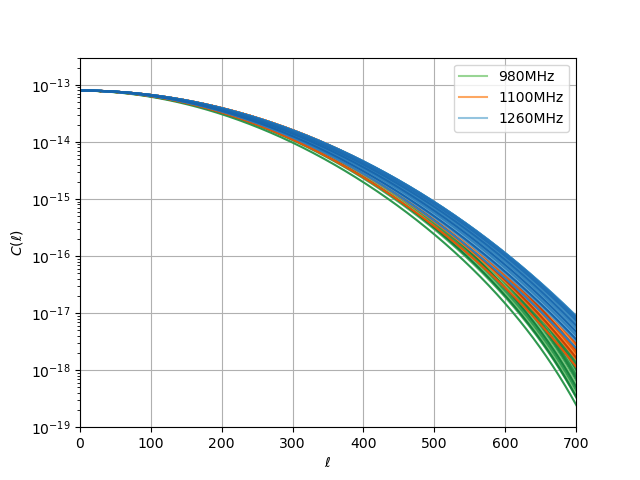}
    \caption{Auto angular power spectra for the \emph{Double-Rectangular} arrangement beams. We computed the spectrum for each one of the 28 horns of the arrangement for the frequencies of 980\,MHz, 1100\,MHz and 1260\,MHz. One can see that low frequencies have lower power at high multipoles. The curves with similar shades of colour correspond to the same frequency but different horns, so the reader can also see the affect due to the location in the optical plane. The intensities were previously normalised with respect to the integral over the simulated area yielding to an adimensional spectrum (with no units), which also represents the window function of the eventual survey.   }
    \label{fig.spectra}
\end{figure}

In order to compute the beams spectra, our first step was convert our data to {\tt HEALPix} format. We linearly interpolated the values of intensity of each beam from the original $u-v$ coordinates to the coordinates of the centers of the pixels of a {\tt HEALPix} pixelated sphere with $N_\text{side} = 1024$. This $N_\text{side}$ value was set in order to make the new resolution compatible with that of our GRASP simulations. We also set to zero the intensity outside the area where the simulations were performed since it is expected to be very low away from the beam center. After this, we compute the projections in harmonic space and all the angular power spectra with a {\tt PseudoPower} code \citep{loureiro2019}. The intensities of each map were previously normalized by its sum over the pixels. The spectra for the beams of the \emph{Double-Rectangular} arrangement are shown in Fig. \ref{fig.spectra} for the frequencies of 980\,MHz, 1100\,MHz and 1260\,MHz. We see that these spectra reach lower values in high-$\ell$ for lower frequencies. In the same figure the reader can infer the effect due to the horn location over the arrangements. This reflects the aberrations of the when displaced away from the centre of the optical plane (\ref{fig.beams12}).  

We conclude that the beams are smooth enough and concentrated enough so as to not produce any significant effects in the measurements of the power spectrum of the 21-cm at the angular scales relevant to BINGO. Furthermore, such modelling can be used in order to produce a more realistic fit to the beams so that this can be used for map-making, although due to the underillumination of the secondary mirror, the beams are very close to Gaussian in the center of the field.

\subsection{Spillover\label{section:spillover}}

The telescope design is set up in such a way that the secondary mirror is underilluminated by the horns given the combination of the focal length chosen and the angular aperture of the horns. As such the spillover of this setup is extremely low at the boresight of the telescope, which is the horn that will point at declination $\delta = -15^{\circ}$ and is located at position $(x,y)=(0,0)$.

\begin{figure}
    \centering
    \includegraphics[scale=0.17]{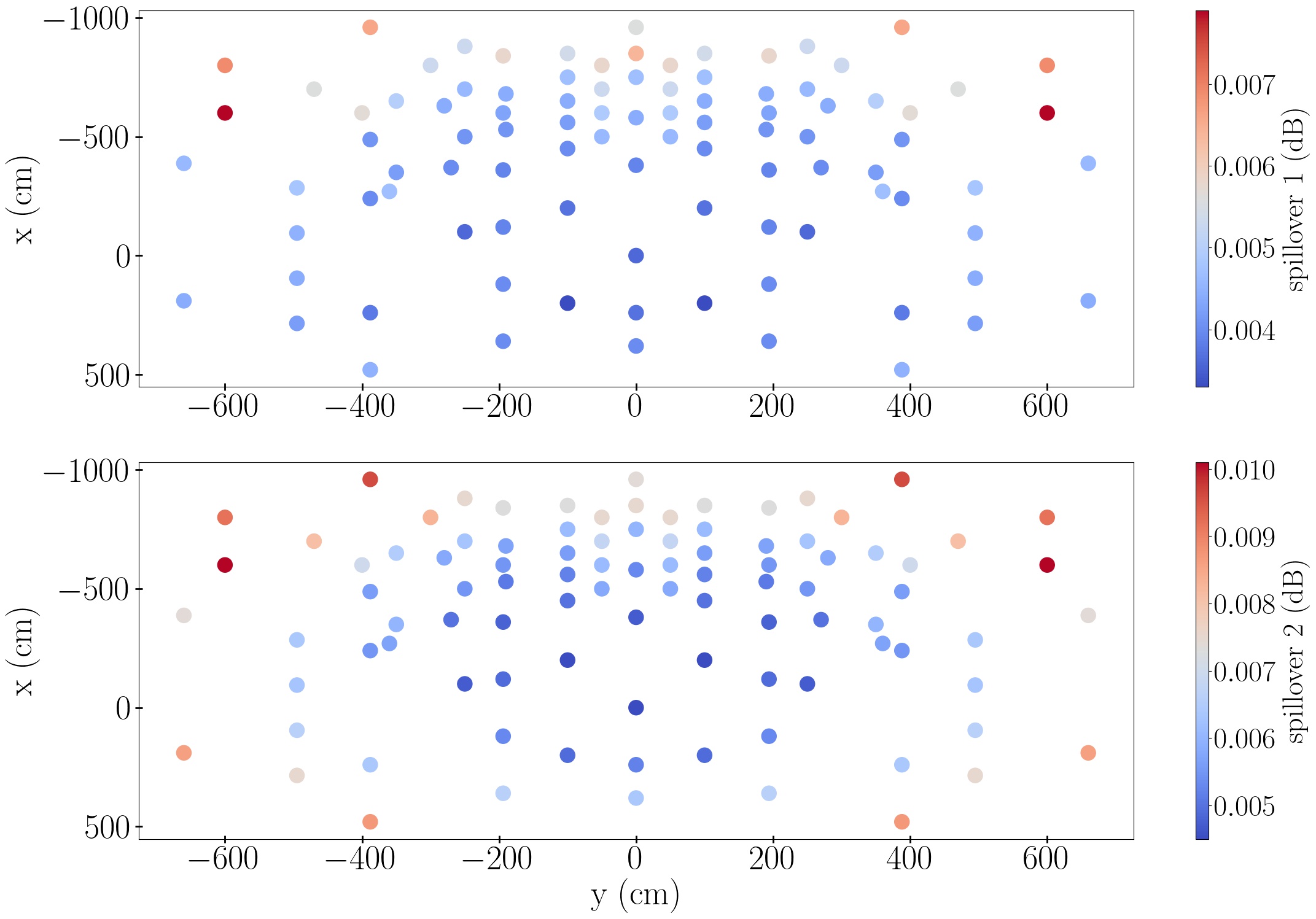}
    \caption{(Top) spillover1 is in relation to the secondary reflector. We can see that at the focus of the telescope the spillover1 is 0.0036\,dB and that at worst it raises to 0.008\,dB at the edges of the focal plane. (Bottom) spillover2 is in relation to the primary reflector. Its  spillover at the center is 0.0045\,dB and the worst is 0.0101\,dB.}
    \label{fig.spillover}    
\end{figure}

We need to assess if the optical arrangement, which requires the horns to point to a given location in the focal plane, has a spillover that is significantly different for each of the horns located outside the main focus of the telescope.  In Fig. \ref{fig.spillover}, we plot the calculated spillover, which is expected given the optical arrangement of the telescope as well as the shape proposed by the focal plane after the optimisation performed in this paper. We can see that at the focus of the telescope the spillover is very low at around 0.003\,dB (lowest value on the scale in Fig.\ref{fig.spillover}). Fig. \ref{fig.spillover} shows us the spillover more than doubles in dB at the edges of the considered focal plane, however even at the locations furthest from the center of the focal plane, it remains at a value lower than 0.007 in dB.

The above results are encouraging as they mean that even without shielding of the main mirror and without any further measures to avoid ground pickup, the spillover for the entire system should remain below 0.007\,dB and in most cases around $0.003-0.004$\,dB. We are considering using a ground shield around the primary and at the bottom of the secondary dish, which should considerably improve the above values.

\section{Conclusions}\label{sec:conclusions}

\alex{This work presents the focal surface calculation as well as a detailed description of the analysis conducted to produce the optical arrangement for the BINGO telescope.}

Our target was to arrive at a cross-Dragone configuration that was under-illuminated, with a very low cross polarisation leakage and sidelobe levels. For each horn, the attenuation obtained from our simulations is better than 60\,dB a few degrees away from the center of the beam, which is crucial to obtain data that is robust to sidelobe contamination, especially for observations performed near the Galactic center.

The polarisation leakage in the Stokes parameters $U$ and $V$ is attenuated by $\approx 25\,dB$ compared to the $I$ intensity in the beam across the field of view. These results indicate that the proposed configuration is appropriate for the scientific goals of BINGO.

We have also analysed the shape and orientation of the focal plane of this cross Dragone configuration. We find a non-planar focal surface which has a complex shape,
mainly due to the fact that multiple local maxima are present when optimising the location, as well as angular orientation of the horns away from the focus of the telescope. This lead us to find specific local maxima of the peak response of the field as a function of the horn positions (most specifically the z direction defined in the frame of the horns) in order to have a suitably shaped focal plane. Given the focal length coupled with the size of the horns, the Nyquist properties of the focal plane were investigated. A suitable shape of the focal surface was found in order to optimise the coverage of the final survey. 

The level of optical aberrations was investigated for detectors away from the focus of the instrument. One of our first commissioning tasks will be to compare the measured beam patterns with the simulated ones. We find that, for the \emph{Double Rectangular} array, the optical aberrations are acceptable and only modify the beam at the level of the sidelobes a couple of degrees away from the main peak beam. Furthermore we find that the attenuation of the main peak of the beam is maintained at reasonable levels, i.e., at the edge of the chosen field the attenuation is at most 0.5\,dB and is maintained within a fraction of a $dB_i$ in most places inside the focal surface.

\textcolor{black}{To maintain a focus tolerance of 0.005 dB, we find that the depth of field for our focal surface should be $\sim 10$ cm. This is in agreement with the mechanical tolerances and requirements of the engineering project designed to host the focal surface, as well as the noise budget from the full BINGO detection system. We have checked that, for several positions within our focal plane, our ellipticity requirement of 0.1 is satisfied and this is indeed the case for the optical design proposed. Furthermore we have checked that the cross Dragone design used in BINGO Phase 1 achieves a polarization purity of -30 dB as demanded by scientific requirements and this is the case for most of the focal plane, as can be seen from Fig.\ref{fig.stokes_param_dec}. The only place where the requirements are close possibly below the simulation is at declination $\delta = -25^{\circ}$,  where the V polarization gets to -25 dBi from the peak response of the I polarization. On the other hand, U and Q Stokes parameters are well below -30 dBi benchmark, as required for polarization purity.}

We arrived at an optimal configuration, named \emph{Double Rectangular}, that meets the requirements for the scientific performance of the instrument. This arrangement, coupled to a system that allows the vertical displacement of the horns in subsequent years of the survey allows for a homogeneous and gap-free coverage of the sky. The simulations indicate a low spillover, good beam performance and, according to the results presented here, the \emph{Double Rectangular} arrangement is considered the nominal configuration for the Phase 1 survey.


\begin{acknowledgements}
The BINGO project is supported by FAPESP grant 2014/07885-0; the support from CNPq is also gratefully acknowledged (E.A.).
F.B.A. acknowledges the UKRI-FAPESP grant 2019/05687-0, and FAPESP and USP for Visiting Professor Fellowships where this work has been developed. 
We thank Edvaldo Barbosa Guimarães Filho for his help and support in relation to codes and optimisations. 
K.S.F.F. would like to thank FAPESP for grant 2017/21570-0. 
C.A.W. acknowledges a CNPq grant 2014/313.597. 
 T.V. acknowledges CNPq Grant 308876/2014-8. 
 E.J.M. acknowledges the support by CAPES.
 A.A.C. acknowledges financial support from the China Postdoctoral Science Foundation, grant number 2020M671611. 
 B. W. and A.A.C. were also supported by the key project of NNSFC under grant 11835009. 
 V.L. acknowledges the postdoctoral FAPESP grant 2018/02026-0.
 C.P.N. thanks S{\~a}o Paulo Research Foundation (FAPESP) for financial support through grant 2019/06040-0. M.P. acknowledges funding from a FAPESP Young Investigator fellowship, grant 2015/19936-1.
R.G.L. thanks CAPES (process 88881.162206/2017-01) and the Alexander von Humboldt Foundation for the financial support. 
A.R.Q., F.A.B., L.B., and M.V.S. acknowledge PRONEX/CNPq/FAPESQ-PB (Grant no. 165/2018).
L. S. is supported by the National Key R\&D Program of China (2020YFC2201600).
J.Z. was supported by IBS under the project code, IBS-R018-D1.
\end{acknowledgements}

%
%






   
  



\bibliographystyle{aa}
\bibliography{bibliog} 

\begin{thebibliography}{38}
\expandafter\ifx\csname natexlab\endcsname\relax\def\natexlab#1{#1}\fi

\bibitem[{Abdalla {et~al.}(2021)}]{2020_project}
Abdalla, E. {et~al.} 2021 [\eprint[arXiv]{2107.01633}]

\bibitem[{{Bandura} {et~al.}(2014){Bandura}, {Addison}, {Amiri}, {Bond},
  {Campbell-Wilson}, {Connor}, {Cliche}, {Davis}, {Deng}, {Denman}, {Dobbs},
  {Fandino}, {Gibbs}, {Gilbert}, {Halpern}, {Hanna}, {Hincks}, {Hinshaw},
  {H{\"o}fer}, {Klages}, {Landecker}, {Masui}, {Mena Parra}, {Newburgh}, {Pen},
  {Peterson}, {Recnik}, {Shaw}, {Sigurdson}, {Sitwell}, {Smecher}, {Smegal},
  {Vanderlinde}, \& {Wiebe}}]{2014SPIE.9145E..22B}
{Bandura}, K., {Addison}, G.~E., {Amiri}, M., {et~al.} 2014, in Society of
  Photo-Optical Instrumentation Engineers (SPIE) Conference Series, Vol. 9145,
  Ground-based and Airborne Telescopes V, 914522

\bibitem[{{Battye} {et~al.}(2012){Battye}, {Brown}, {Browne}, {Davis},
  {Dewdney}, {Dickinson}, {Heron}, {Maffei}, {Pourtsidou}, \&
  {Wilkinson}}]{2012arXiv1209.1041B}
{Battye}, R.~A., {Brown}, M.~L., {Browne}, I.~W.~A., {et~al.} 2012, arXiv
  e-prints, arXiv:1209.1041

\bibitem[{{Battye} {et~al.}(2013){Battye}, {Browne}, {Dickinson}, {Heron},
  {Maffei}, \& {Pourtsidou}}]{2013MNRAS.434.1239B}
{Battye}, R.~A., {Browne}, I.~W.~A., {Dickinson}, C., {et~al.} 2013, \mnras,
  434, 1239

\bibitem[{{Battye} {et~al.}(2004){Battye}, {Davies}, \&
  {Weller}}]{2004MNRAS.355.1339B}
{Battye}, R.~A., {Davies}, R.~D., \& {Weller}, J. 2004, \mnras, 355, 1339

\bibitem[{{Bautista} {et~al.}(2020){Bautista}, {Paviot}, {Maga{\~n}a}, {de la
  Torre}, {Fromenteau}, {Gil-Mar{\'\i}n}, {Ross}, {Burtin}, {Dawson}, {Hou},
  {Kneib}, {de Mattia}, {Percival}, {Rossi}, {Tojeiro}, {Zhao}, {Zhao}, {Alam},
  {Brownstein}, {Chapman}, {Choi}, {Chuang}, {Escoffier}, {de la Macorra}, {du
  Mas des Bourboux}, {Mohammad}, {Moon}, {M{\"u}ller}, {Nadathur}, {Newman},
  {Schneider}, {Seo}, \& {Wang}}]{2020MNRAS.tmp.2651B}
{Bautista}, J.~E., {Paviot}, R., {Maga{\~n}a}, M.~V., {et~al.} 2020, \mnras
  [\eprint[arXiv]{2007.08993}]

\bibitem[{{Beutler} {et~al.}(2011){Beutler}, {Blake}, {Colless}, {Jones},
  {Staveley-Smith}, {Campbell}, {Parker}, {Saunders}, \&
  {Watson}}]{2011MNRAS.416.3017B}
{Beutler}, F., {Blake}, C., {Colless}, M., {et~al.} 2011, \mnras, 416, 3017

\bibitem[{{Blake} {et~al.}(2011){Blake}, {Davis}, {Poole}, {Parkinson},
  {Brough}, {Colless}, {Contreras}, {Couch}, {Croom}, {Drinkwater}, {Forster},
  {Gilbank}, {Gladders}, {Glazebrook}, {Jelliffe}, {Jurek}, {Li}, {Madore},
  {Martin}, {Pimbblet}, {Pracy}, {Sharp}, {Wisnioski}, {Woods}, {Wyder}, \&
  {Yee}}]{2011MNRAS.415.2892B}
{Blake}, C., {Davis}, T., {Poole}, G.~B., {et~al.} 2011, \mnras, 415, 2892

\bibitem[{{Chen}(2012)}]{2012IJMPS..12..256C}
{Chen}, X. 2012, in International Journal of Modern Physics Conference Series,
  Vol.~12, International Journal of Modern Physics Conference Series, 256--263

\bibitem[{Costa {et~al.}(2021)}]{2020_forecast}
Costa, A.~A. {et~al.} 2021 [\eprint[arXiv]{2107.01639}]

\bibitem[{{DeBoer} {et~al.}(2017){DeBoer}, {Parsons}, {Aguirre}, {Alexander},
  {Ali}, {Beardsley}, {Bernardi}, {Bowman}, {Bradley}, {Carilli}, {Cheng}, {de
  Lera Acedo}, {Dillon}, {Ewall-Wice}, {Fadana}, {Fagnoni}, {Fritz},
  {Furlanetto}, {Glendenning}, {Greig}, {Grobbelaar}, {Hazelton}, {Hewitt},
  {Hickish}, {Jacobs}, {Julius}, {Kariseb}, {Kohn}, {Lekalake}, {Liu}, {Loots},
  {MacMahon}, {Malan}, {Malgas}, {Maree}, {Martinot}, {Mathison}, {Matsetela},
  {Mesinger}, {Morales}, {Neben}, {Patra}, {Pieterse}, {Pober}, {Razavi-Ghods},
  {Ringuette}, {Robnett}, {Rosie}, {Sell}, {Smith}, {Syce}, {Tegmark},
  {Thyagarajan}, {Williams}, \& {Zheng}}]{2017PASP..129d5001D}
{DeBoer}, D.~R., {Parsons}, A.~R., {Aguirre}, J.~E., {et~al.} 2017, \pasp, 129,
  045001

\bibitem[{{Dragone}(1978)}]{1978ATTTJ..57.2663D}
{Dragone}, C. 1978, AT T Technical Journal, 57, 2663

\bibitem[{{du Mas des Bourboux} {et~al.}(2020){du Mas des Bourboux}, {Rich},
  {Font-Ribera}, {de Sainte Agathe}, {Farr}, {Etourneau}, {Le Goff}, {Cuceu},
  {Balland}, {Bautista}, {Blomqvist}, {Brinkmann}, {Brownstein}, {Chabanier},
  {Chaussidon}, {Dawson}, {Gonz{\'a}lez-Morales}, {Guy}, {Lyke}, {de la
  Macorra}, {Mueller}, {Myers}, {Nitschelm}, {Mu{\~n}oz Guti{\'e}rrez},
  {Palanque-Delabrouille}, {Parker}, {Percival}, {P{\'e}rez-R{\`a}fols},
  {Petitjean}, {Pieri}, {Ravoux}, {Rossi}, {Schneider}, {Seo}, {Slosar},
  {Stermer}, {Vivek}, {Y{\`e}che}, \& {Youles}}]{2020ApJ...901..153D}
{du Mas des Bourboux}, H., {Rich}, J., {Font-Ribera}, A., {et~al.} 2020, \apj,
  901, 153

\bibitem[{{Eisenstein} {et~al.}(2005){Eisenstein}, {Zehavi}, {Hogg},
  {Scoccimarro}, {Blanton}, {Nichol}, {Scranton}, {Seo}, {Tegmark}, {Zheng},
  {Anderson}, {Annis}, {Bahcall}, {Brinkmann}, {Burles}, {Castand er},
  {Connolly}, {Csabai}, {Doi}, {Fukugita}, {Frieman}, {Glazebrook}, {Gunn},
  {Hendry}, {Hennessy}, {Ivezi{\'c}}, {Kent}, {Knapp}, {Lin}, {Loh}, {Lupton},
  {Margon}, {McKay}, {Meiksin}, {Munn}, {Pope}, {Richmond}, {Schlegel},
  {Schneider}, {Shimasaku}, {Stoughton}, {Strauss}, {SubbaRao}, {Szalay},
  {Szapudi}, {Tucker}, {Yanny}, \& {York}}]{2005ApJ...633..560E}
{Eisenstein}, D.~J., {Zehavi}, I., {Hogg}, D.~W., {et~al.} 2005, \apj, 633, 560

\bibitem[{Fornazier {et~al.}(2021)}]{2020_component_separation}
Fornazier, K. S.~F. {et~al.} 2021 [\eprint[arXiv]{2107.01637}]

\bibitem[{{Fruchter} \& {Hook}(2002)}]{2002PASP..114..144F}
{Fruchter}, A.~S. \& {Hook}, R.~N. 2002, \pasp, 114, 144

\bibitem[{{Gong} {et~al.}(2012){Gong}, {Cooray}, {Silva}, {Santos}, {Bock},
  {Bradford}, \& {Zemcov}}]{2012ApJ...745...49G}
{Gong}, Y., {Cooray}, A., {Silva}, M., {et~al.} 2012, \apj, 745, 49

\bibitem[{{Harper} {et~al.}(2018){Harper}, {Dickinson}, {Battye},
  {Roychowdhury}, {Browne}, {Ma}, {Olivari}, \& {Chen}}]{1f_HI_2018}
{Harper}, S.~E., {Dickinson}, C., {Battye}, R.~A., {et~al.} 2018, \mnras, 478,
  2416

\bibitem[{{Levi} {et~al.}(2019){Levi}, {Allen}, {Raichoor}, {Baltay}, {BenZvi},
  {Beutler}, {Bolton}, {Castander}, {Chuang}, {Cooper}, {Cuby}, {Dey},
  {Eisenstein}, {Fan}, {Flaugher}, {Frenk}, {Gonzalez-Morales}, {Graur}, {Guy},
  {Habib}, {Honscheid}, {Juneau}, {Kneib}, {Lahav}, {Lang}, {Leauthaud},
  {Lusso}, {de la Macorra}, {Manera}, {Martini}, {Mao}, {Newman},
  {Palanque-Delabrouille}, {Percival}, {Prieto}, {Rockosi}, {Ruhlmann-Kleider},
  {Schlegel}, {Seo}, {Song}, {Tarle}, {Wechsler}, {Weinberg}, {Yeche}, \&
  {Zu}}]{2019BAAS...51g..57L}
{Levi}, M., {Allen}, L.~E., {Raichoor}, A., {et~al.} 2019, in Bulletin of the
  American Astronomical Society, Vol.~51, 57

\bibitem[{Liccardo {et~al.}(2021)}]{2020_sky_simulation}
Liccardo, V. {et~al.} 2021 [\eprint[arXiv]{2107.01636}]

\bibitem[{{Lidz} {et~al.}(2011){Lidz}, {Furlanetto}, {Oh}, {Aguirre}, {Chang},
  {Dor{\'e}}, \& {Pritchard}}]{2011ApJ...741...70L}
{Lidz}, A., {Furlanetto}, S.~R., {Oh}, S.~P., {et~al.} 2011, \apj, 741, 70

\bibitem[{{Loureiro} {et~al.}(2019){Loureiro}, {Moraes}, {Abdalla}, {Cuceu},
  {McLeod}, {Whiteway}, {Balan}, {Benoit-L{\'e}vy}, {Lahav}, {Manera},
  {Rollins}, \& {Xavier}}]{loureiro2019}
{Loureiro}, A., {Moraes}, B., {Abdalla}, F.~B., {et~al.} 2019, \mnras, 485, 326

\bibitem[{{Madau} {et~al.}(1997){Madau}, {Meiksin}, \&
  {Rees}}]{1997ApJ...475..429M}
{Madau}, P., {Meiksin}, A., \& {Rees}, M.~J. 1997, \apj, 475, 429

\bibitem[{{Masui} {et~al.}(2013){Masui}, {Switzer}, {Banavar}, {Bandura},
  {Blake}, {Calin}, {Chang}, {Chen}, {Li}, {Liao}, {Natarajan}, {Pen},
  {Peterson}, {Shaw}, \& {Voytek}}]{GBTcross2013}
{Masui}, K.~W., {Switzer}, E.~R., {Banavar}, N., {et~al.} 2013, \apjl, 763, L20

\bibitem[{{Newburgh} {et~al.}(2016){Newburgh}, {Bandura}, {Bucher}, {Chang},
  {Chiang}, {Cliche}, {Dav{\'e}}, {Dobbs}, {Clarkson}, {Ganga}, {Gogo},
  {Gumba}, {Gupta}, {Hilton}, {Johnstone}, {Karastergiou}, {Kunz}, {Lokhorst},
  {Maartens}, {Macpherson}, {Mdlalose}, {Moodley}, {Ngwenya}, {Parra},
  {Peterson}, {Recnik}, {Saliwanchik}, {Santos}, {Sievers}, {Smirnov},
  {Stronkhorst}, {Taylor}, {Vanderlinde}, {Van Vuuren}, {Weltman}, \&
  {Witzemann}}]{2016SPIE.9906E..5XN}
{Newburgh}, L.~B., {Bandura}, K., {Bucher}, M.~A., {et~al.} 2016, in Society of
  Photo-Optical Instrumentation Engineers (SPIE) Conference Series, Vol. 9906,
  Ground-based and Airborne Telescopes VI, ed. H.~J. {Hall}, R.~{Gilmozzi}, \&
  H.~K. {Marshall}, 99065X

\bibitem[{{Peel} {et~al.}(2019){Peel}, {Wuensche}, {Abdalla}, {Ant{\'o}n},
  {Barosi}, {Browne}, {Caldas}, {Dickinson}, {Fornazier}, {Monstein},
  {Strauss}, {Tancredi}, \& {Villela}}]{Peel:2019}
{Peel}, M.~W., {Wuensche}, C.~A., {Abdalla}, E., {et~al.} 2019, Journal of
  Astronomical Instrumentation, 8, 1940005

\bibitem[{{Peterson} {et~al.}(2009){Peterson}, {Aleksan}, {Ansari}, {Band ura},
  {Bond}, {Bunton}, {Carlson}, {Chang}, {DeJongh}, {Dobbs}, {Dodelson},
  {Darhmaoui}, {Gnedin}, {Halpern}, {Hogan}, {Le Goff}, {Liu}, {Legrouri},
  {Loeb}, {Loudiyi}, {Magneville}, {Marriner}, {McGinnis}, {McWilliams},
  {Moniez}, {Palanque-Delabruille}, {Pasquinelli}, {Pen}, {Rich}, {Scarpine},
  {Seo}, {Sigurdson}, {Seljak}, {Stebbins}, {Steffen}, {Stoughton}, {Timbie},
  {Vallinotto}, \& {Teche}}]{2009astro2010S.234P}
{Peterson}, J.~B., {Aleksan}, R., {Ansari}, R., {et~al.} 2009, in astro2010:
  The Astronomy and Astrophysics Decadal Survey, Vol. 2010, 234

\bibitem[{{Poidevin} {et~al.}(2018){Poidevin}, {Rubino-Martin},
  {Genova-Santos}, {Rebolo}, {Aguiar}, {Gomez-Renasco}, {Guidi.}, {Gutierrez},
  {Hoyland}, {Lopez-Caraballo}, {Oria Carreras}, {Pelaez-Santos},
  {Perez-De-Taoro}, {Ruiz-Granados}, {Tramonte}, {Vega-Moreno},
  {Viera-Curbelo}, {Vignaga}, {Martinez-Gonzalez}, {Barreiro}, {Casaponsa},
  {Casas}, {Diego}, {Fernandez-Cobos}, {Herranz}, {Lopez-Caniego}, {Ortiz},
  {Vielva}, {Artal}, {Aja}, {Cagigas}, {Cano}, {De La Fuente}, {Mediavilla},
  {Teran}, {Villa}, {Piccirillo}, {Dickinson}, {Grainge}, {Harper},
  {Mcculloch}, {Melhuish}, {Pisano}, {Watson}, {Lasenby}, {Ashdown}, {Perrott},
  {Razavi-Ghods}, {Titterington}, \& {Scott}}]{QUIJOTE2018}
{Poidevin}, F., {Rubino-Martin}, J.~A., {Genova-Santos}, R., {et~al.} 2018,
  arXiv e-prints, arXiv:1802.04594

\bibitem[{{Pullen} {et~al.}(2014){Pullen}, {Dor{\'e}}, \&
  {Bock}}]{2014ApJ...786..111P}
{Pullen}, A.~R., {Dor{\'e}}, O., \& {Bock}, J. 2014, \apj, 786, 111

\bibitem[{{Santos} {et~al.}(2017){Santos}, {Cluver}, {Hilton}, {Jarvis},
  {Jozsa}, {Leeuw}, {Smirnov}, {Taylor}, {Abdalla}, {Afonso}, {Alonso},
  {Bacon}, {Bassett}, {Bernardi}, {Bull}, {Camera}, {Chiang}, {Colafrancesco},
  {Ferreira}, {Fonseca}, {van der Heyden}, {Heywood}, {Knowles}, {Lochner},
  {Ma}, {Maartens}, {Makhathini}, {Moodley}, {Pourtsidou}, {Prescott},
  {Sievers}, {Spekkens}, {Vaccari}, {Weltman}, {Whittam}, {Witzemann}, {Wolz},
  \& {Zwart}}]{2017arXiv170906099S}
{Santos}, M.~G., {Cluver}, M., {Hilton}, M., {et~al.} 2017, arXiv e-prints,
  arXiv:1709.06099

\bibitem[{{Square Kilometre Array Cosmology Science Working Group}
  {et~al.}(2020){Square Kilometre Array Cosmology Science Working Group},
  {Bacon}, {Battye}, {Bull}, {Camera}, {Ferreira}, {Harrison}, {Parkinson},
  {Pourtsidou}, {Santos}, {Wolz}, {Abdalla}, {Akrami}, {Alonso},
  {Andrianomena}, {Ballardini}, {Bernal}, {Bertacca}, {Bengaly}, {Bonaldi},
  {Bonvin}, {Brown}, {Chapman}, {Chen}, {Chen}, {Cunnington}, {Davis},
  {Dickinson}, {Fonseca}, {Grainge}, {Harper}, {Jarvis}, {Maartens}, {Maddox},
  {Padmanabhan}, {Pritchard}, {Raccanelli}, {Rivi}, {Roychowdhury},
  {Sahl{\'e}n}, {Schwarz}, {Siewert}, {Viel}, {Villaescusa-Navarro}, {Xu},
  {Yamauchi}, \& {Zuntz}}]{2020PASA...37....7S}
{Square Kilometre Array Cosmology Science Working Group}, {Bacon}, D.~J.,
  {Battye}, R.~A., {et~al.} 2020, \pasa, 37, e007

\bibitem[{{Switzer} {et~al.}(2013){Switzer}, {Masui}, {Bandura}, {Calin},
  {Chang}, {Chen}, {Li}, {Liao}, {Natarajan}, {Pen}, {Peterson}, {Shaw}, \&
  {Voytek}}]{2013MNRAS.434L..46S}
{Switzer}, E.~R., {Masui}, K.~W., {Bandura}, K., {et~al.} 2013, \mnras, 434,
  L46

\bibitem[{{Tran} {et~al.}(2008){Tran}, {Lee}, {Hanany}, {Milligan}, \&
  {Renbarger}}]{2008ApOpt..47..103T}
{Tran}, H., {Lee}, A., {Hanany}, S., {Milligan}, M., \& {Renbarger}, T. 2008,
  \ao, 47, 103

\bibitem[{{Wolz} {et~al.}(2015){Wolz}, {Blake}, {Abdalla}, {Anderson}, {Chang},
  {Li}, {Masui}, {Switzer}, {Pen}, {Voytek}, \& {Yadav}}]{GBTcross2015}
{Wolz}, L., {Blake}, C., {Abdalla}, F.~B., {et~al.} 2015, arXiv e-prints,
  arXiv:1510.05453

\bibitem[{{Wolz} {et~al.}(2021){Wolz}, {Pourtsidou}, {Masui}, {Chang},
  {Bautista}, {Mueller}, {Avila}, {Bacon}, {Percival}, {Cunnington},
  {Anderson}, {Chen}, {Kneib}, {Li}, {Liao}, {Pen}, {Peterson}, {Rossi},
  {Schneider}, {Yadav}, \& {Zhao}}]{GBTcross2021}
{Wolz}, L., {Pourtsidou}, A., {Masui}, K.~W., {et~al.} 2021, arXiv e-prints,
  arXiv:2102.04946

\bibitem[{{Wuensche} {et~al.}(2020){Wuensche}, {Reitano}, {Peel}, {Browne},
  {Maffei}, {Abdalla}, {Radcliffe}, {Abdalla}, {Barosi}, {Liccardo}, {Mericia},
  {Pisano}, {Strauss}, {Vieira}, {Villela}, \& {Wang}}]{2020ExA....50..125W}
{Wuensche}, C.~A., {Reitano}, L., {Peel}, M.~W., {et~al.} 2020, Experimental
  Astronomy, 50, 125

\bibitem[{Wuensche {et~al.}(2021)}]{2020_BINGO_Instrument}
Wuensche, C.~A. {et~al.} 2021 [\eprint[arXiv]{2107.01634}]

\bibitem[{Zhang {et~al.}(2021)}]{2020_mock_simulations}
Zhang, J. {et~al.} 2021 [\eprint[arXiv]{2107.01638}]

\end{thebibliography}

\begin{appendix} 

\section{\textcolor{black}{GRASP setup} \label{AppendixC}}
\textcolor{black}{We can divide the GRASP calculation process used into four steps: (1) obtain currents (\emph{Get Current}) with a specific feed as source and the sub-reflector (hyperbolic dish) as target, (2) obtain currents (\emph{Get Current}) using sub-reflector as source and main reflector (parabolic dish) as target, (3) calculate the field (\emph{Get Field}) using the main reflector as source, and (4) calculate the field (\emph{Add Field}) using a feed as source. The currents calculation using \emph{PO Analysis}\footnote{PO = Physical Optics} also use PTD\footnote{PTD = Physical Theory of Diffraction} correction, that is, we used PO+PTD currents. PTD corresponding to currents from the edges of the reflector. The PO+PTD need of three parameters (po1, po2 and ptd) to calculate to convergence of the currents calculations and provided by auto-convergence (established \textbf{ON}). Those parameters are estimated by the geometry of the optical system and by the wavelength used. It can be seen in tables \ref{tab:task1} and \ref{tab:task2} examples of some values of the parameters. In that cases we chose 5 different positions on focal plane to exemplify the values. The field accuracy used was the GRASP standard value: -80 dB in relation to the peak. Third and Fourth steps build the spherical grid file in calculating the field. Spherical grid files (section 4) was configured in uv grid, both x-range and y-range with: start = -0.5, end = 0.5 and np = 151\footnote{\textcolor{black}{To use partial sky does not change spillover values and neither respective intensities, as long as the beam in the sky is covered. What can change the intensity is its accuracy when changing np values.}}. Also, rectangular truncation and linear polarization. For 5 cases in \ref{tab:task3} and \ref{tab:task4}, it can be seen parameter values (for integration) to obtain the field from main reflector (\ref{tab:task3}) and from a feed (\ref{tab:task4}).}
\begin{table*}[h]
    \centering
    \scriptsize
    \begin{tabular}{ |P{0.6cm}|| P{0.7cm}|P{0.7cm}|P{0.7cm}|P{0.7cm}|P{0.9cm}|P{0.7cm}|P{0.7cm}|P{0.7cm}|P{1.cm}|P{1.cm}|}
    \hline
    \multicolumn{11}{|c|}{Task 1 - Get Currents} \\
    \hline
     &  x (cm) & y (cm) & z (cm) & 	$\theta$ (deg) & $\phi$ (deg) & po1 & po2 & ptd & po target & ptd target\\
    \hline
P0 &    0 &    0 &    0 & 0.00 & 0.00 & 170 & 570 & 570 & 62350 & 570\\
P1 & -990 &  305 &   98 & 9.28 & -11.19 & 185 & 640 & 640 & 76082 & 640 \\
P2 &  510 & -305 &   45 & 4.08 & 145.08 & 160 & 540 & 530 & 55616 & 530\\
P3 & -930 & -305 &  102 & 8.98 & 15.77 & 185 & 630 & 630 & 74918 & 630\\
P4 &  450 &  305 &   36 & 3.62 & -149.32 & 165 & 540 & 540 & 57365 & 540\\ 
    \hline
    \end{tabular}
\caption{\textcolor{black}{We present the principle parameters for convergence information in current calculations when a feed is source and sub-reflector is target for 5 different positions on focal plane (P0, P1, P2, P3 and P4). It can be see that the position on focal plane influences parameter values. Positions away from the ground (smaller x value) require higher values.}}
\label{tab:task1}
\end{table*}

\begin{table*}[h]
    \centering
    \scriptsize
    \begin{tabular}{ |P{0.6cm}|| P{0.7cm}|P{0.7cm}|P{0.7cm}|P{0.7cm}|P{0.8cm}|P{0.7cm}|P{0.7cm}|P{0.7cm}|P{1.cm}|P{1.cm}|P{1.1cm}|}
    \hline
    \multicolumn{12}{|c|}{Task 2 - Get Currents} \\
    \hline
& x (cm) & y (cm)  & z (cm)  & $\theta$ (deg) & $\phi$ (deg) & po1 & po2 & ptd & po target & po source & ptd source\\
\hline
P0 &    0 &    0 &    0 & -0.00 & 0.00 & 305 & 940 & 940 & 183690 & 62350 & 570\\
P1 & -990 &  305 &   98 & 9.28 & -11.19 & 305 & 900 & 860 & 175929 & 76082 & 640\\
P2 &  510 & -305 &   45 & 4.08 & 145.08 & 305 & 980 & 980 & 191462 & 55616 & 530\\
P3 & -930 & -305 &  102 & 8.98 & 15.77 & 305 & 900 & 860 & 175929 & 74918 & 630\\
P4 &  450 &  305 &   36 & 3.62 & -149.32 & 305 & 980 & 980 & 191462 & 57365 & 540\\
    \hline
    \end{tabular}
\caption{\textcolor{black}{We present the principal parameters for convergence information in current calculations when a sub-reflector is source and main reflector is target for 5 different positions on the focal plane (P0, P1, P2, P3 and P4).}}
\label{tab:task2}
\end{table*}

\begin{table*}[h]
    \centering
    \scriptsize
    \begin{tabular}{ |P{0.6cm}|| P{0.7cm}|P{0.7cm}|P{0.7cm}|P{0.7cm}|P{0.8cm}|P{1.2cm}|P{1.cm}|P{1.1cm}|}
    \hline
    \multicolumn{9}{|c|}{Task 3 - Get Fields} \\
    \hline
 & x & y & z & $\theta$ & $\phi$ & field points & po source & ptd source\\
\hline
P0 &    0 &    0 &    0 & -0.00 & 0.00 & 22801 & 183690 & 940\\
P1 & -990 &  305 &   98 & 9.28 & -11.19 & 22801 & 175929 & 860\\
P2 &  510 & -305 &   45 & 4.08 & 145.08 & 22801 & 191462 & 980\\
P3 & -930 & -305 &  102 & 8.98 & 15.77 & 22801 & 175929 & 860\\
P4 &  450 &  305 &   36 & 3.62 & -149.32 & 22801 & 191462 & 980\\
    \hline
    \end{tabular}
\caption{\textcolor{black}{We present the principal parameters to calculate the field from main reflector for 5 different positions on the focal plane (P0, P1, P2, P3 and P4).}}
\label{tab:task3}
\end{table*}

\begin{table*}[h]
    \centering
    \scriptsize
    \begin{tabular}{ |P{0.6cm}|| P{0.7cm}|P{0.7cm}|P{0.7cm}|P{0.7cm}|P{0.8cm}|P{1.2cm}|P{1.cm}|P{1.1cm}|}
    \hline
    \multicolumn{9}{|c|}{Task 4 - Add Fields} \\
\hline
 & x & y & z & $\theta$ & $\phi$ & field points & po source & ptd source\\
\hline
P0 &    0 &    0 &    0 & -0.00 & 0.00 & 22801 & 62350 & 570\\
P1 & -990 &  305 &   98 & 9.28 & -11.19 & 22801 & 76082 & 640\\
P2 &  510 & -305 &   45 & 4.08 & 145.08 & 22801 & 55616 & 530\\
P3 & -930 & -305 &  102 & 8.98 & 15.77 & 22801 & 74918 & 630\\
P4 &  450 &  305 &   36 & 3.62 & -149.32 & 22801 & 57365 & 540\\
    \hline
    \end{tabular}
\caption{\textcolor{black}{We present the principal parameters to calculate the field from a feed for 5 different positions on the focal plane (P0, P1, P2, P3 and P4).}}
\label{tab:task4}
\end{table*}

\section{Calibrated horns parameters\label{AppendixA} }
We chose random positions on the focal plane to able to obtain the fit for parameters z, $\theta$ and $\phi$ from x and y coordinates. We using the symmetries of the focal plane to obtain more calibrated values. The lines with Modified equal True are the positions where we used the symmetry from $\phi$, as previously described. In Table \ref{tab:general_table} we show all values used in this paper.

\begin{table*}[h]
    \centering
    \scriptsize
    \begin{tabular}{ |P{1.cm}|P{1.cm}|P{1.2cm}|P{1.2cm}|P{1.5cm}|P{1.6
    cm}|P{2.cm}| }
    \hline
    \multicolumn{7}{|c|}{Calibrated positions} \\
    \hline
    x (cm) & y (cm) & z (cm) & $\theta$ (deg) & $\phi$ (deg) & Amplitude (dB) & Modified\\
    \hline
$-960$ & $-388$ & $105.5^{\tiny{+0.1}}_{\tiny{-1.3}} $&$ 9.37^{\tiny{+0.25}}_{\tiny{-0.02}}$ & $26.30^{\tiny{+1.21}}_{\tiny{-0.96}}$ & $50.47$ & True\\[2.5pt] 
$-960$ & 0 & $108.4^{\tiny{+0.4}}_{\tiny{-0.7}}$ & $9.00^{\tiny{+0.20}}_{\tiny{-0.10}}$ & $0.00^{\tiny{+2.63}}_{\tiny{-2.63}} $ & 50.63 & False\\[2.5pt]  
$-840$ & $-194$ & $103.9^{\tiny{+0.8}}_{\tiny{-0.5}}$ & $7.88^{\tiny{+0.13}}_{\tiny{-0.18}}$ & $5.50^{\tiny{+2.40}}_{\tiny{-6.52}}$ & 50.74 & True\\[2.5pt]  
$-600$ & $-194$ & $80.3^{\tiny{+2.4}}_{\tiny{-2.7}}$ & $6.12^{\tiny{+0.53}}_{\tiny{-0.56}}$ & $16.30^{\tiny{+1.07}}_{\tiny{-1.10}}$ & 51.04 & True\\[2.5pt]  
$-600$ & 194 & $80.3^{\tiny{+2.4}}_{\tiny{-2.7}}$ & $6.12^{\tiny{+0.53}}_{\tiny{-0.56}}$ & $-16.30^{\tiny{+1.10}}_{\tiny{-1.07}}$ & 51.04 & True\\[2.5pt]  
$-488$ & $-388$ & $85.2^{\tiny{+0.7}}_{\tiny{-1.5}}$ & $5.70^{\tiny{+0.87}}_{\tiny{-0.22}}$ & $38.00^{\tiny{+7.10}}_{\tiny{-2.47}}$ & 51.22 & True\\[2.5pt]  
$-360$ & 194 & $45.8^{\tiny{+1.5}}_{\tiny{-0.2}}$ & $3.85^{\tiny{+0.58}}_{\tiny{-1.02}}$ & $-26.50^{\tiny{+1.71}}_{\tiny{-6.77}}$ & 51.18 & False\\[2.5pt]  
$-240$ & $-388$ & $24.0^{\tiny{+10.7}}_{\tiny{-11.7}}$ & $4.00^{\tiny{+0.61}}_{\tiny{-0.59}}$ & $58.50^{\tiny{+8.41}}_{\tiny{-9.34}}$ & 51.35 & False\\[2.5pt]  
$-120$ & 194 & $-11.0^{\tiny{+6.6}}_{\tiny{-6.4}}$ & $2.00^{\tiny{+0.32}}_{\tiny{-0.32}}$ & $-55.00^{\tiny{+9.97}}_{\tiny{-8.20}}$ & 51.48 & False\\[2.5pt]  
0 & 0 & $0.0^{\tiny{+18.8}}_{\tiny{-8.5}}$ & $0.00^{\tiny{+0.75}}_{\tiny{-0.45}}$ & $0.00^{\tiny{+0.00}}_{\tiny{-0.00}}$ & 51.64 & False\\[2.5pt]  
120 & -194 & $-10.0^{\tiny{+14.2}}_{\tiny{-12.6}}$ & $1.60^{\tiny{+0.67}}_{\tiny{-0.61}}$ & $100.00^{\tiny{+41.17}}_{\tiny{-12.11}}$ & 51.47 & False\\[2.5pt]  
240 & $-388$ & $35.0^{\tiny{+2.5}}_{\tiny{-2.0}} $ & $3.30^{\tiny{+0.18}}_{\tiny{-0.07}} $ & $117.00^{\tiny{+1.12}}_{\tiny{-3.37}}$ & 51.45 & False\\[2.5pt]  
240 & 0 & $0.0^{\tiny{+19.4}}_{\tiny{-7.1}}$ & $1.50^{\tiny{+0.57}}_{\tiny{-0.52}}$ & $180.00^{\tiny{+20.96}}_{\tiny{-20.96}}$ & 51.60 & False\\[2.5pt]  
360 & 194 & $ 0.0^{\tiny{+9.2}}_{\tiny{-8.7}}$ & $2.70^{\tiny{+0.33}}_{\tiny{-0.52}}$ & $-151.00^{\tiny{+10.12}}_{\tiny{-8.83}}$ & 51.39 & False\\[2.5pt]  
480 & $-388$ & $49.0^{\tiny{+4.7}}_{\tiny{-5.5}}$ & $4.18^{\tiny{+0.27}}_{\tiny{-0.22}}$ & $134.70^{\tiny{+3.61}}_{\tiny{-3.49}}$ & 51.32 & False\\[2.5pt]  
$-380$ & 0 & $46.6^{\tiny{+1.9}}_{\tiny{-4.4}} $ & $3.70^{\tiny{+0.99}}_{\tiny{-0.46}}$ & $0.00^{\tiny{+10.16}}_{\tiny{-10.16}}$ & 51.26 & True\\[2.5pt]  
380 & 0 & $0.0^{\tiny{+10.3}}_{\tiny{-10.2}}$ & $2.40^{\tiny{+0.49}}_{\tiny{-0.45}}$ & $180.00^{\tiny{+0.00}}_{\tiny{-0.00}}$ & 51.53 & False\\[2.5pt]  
$-285$ & 495 & $7.4^{\tiny{+0.6}}_{\tiny{-4.4}}$ & $4.75^{\tiny{+0.70}}_{\tiny{-0.71}}$ & $-55.70^{\tiny{+8.89}}_{\tiny{-8.95}}$ & 51.40 & True\\[2.5pt]  
$-95$ & 495 & $8.2^{\tiny{+9.8}}_{\tiny{-0.7}}$ & $4.00^{\tiny{+0.22}}_{\tiny{-0.24}}$ & $-72.97^{\tiny{+3.36}}_{\tiny{-3.45}}$ & 51.44 & True\\[2.5pt]  
95 & $-495 $ & $14.0^{\tiny{+14.4}}_{\tiny{-13.6}}$ & $3.80^{\tiny{+0.70}}_{\tiny{-0.72}} $ & $93.00^{\tiny{+11.27}}_{\tiny{-11.08}}$ & 51.38 & False\\[2.5pt]  
285 & 495 & $ 52.0^{\tiny{+0.3}}_{\tiny{-1.9}}$ & $4.10^{\tiny{+0.34}}_{\tiny{-0.26}}$ & $-114.00^{\tiny{+13.38}}_{\tiny{-1.33}}$ & 51.20 & True\\[2.5pt]  
190& $-660$ & $62.0^{\tiny{+6.0}}_{\tiny{-6.3}}$  & $5.20^{\tiny{+0.37}}_{\tiny{-0.29}}$  & $104.00^{\tiny{+3.45}}_{\tiny{-4.22}} $ & 51.21 & False\\[2.5pt]  
$-388$ & 660 & $75.0^{\tiny{+ 0.7}}_{\tiny{-2.4}}$ & $6.60^{\tiny{+0.84}}_{\tiny{-0.16}}$ & $-57.35^{\tiny{+0.86}}_{\tiny{-11.49}}$ & 51.02& False\\[2.5pt]  
$-800$ &$ -300$ & $105.0^{\tiny{+9.5}}_{\tiny{-3.2}}$ & $8.70^{\tiny{+0.11}}_{\tiny{-0.41}}$ & $25.30^{\tiny{+2.48}}_{\tiny{-2.76}} $ & 50.92 & False\\[2.5pt]  
$-880$ & $-250$ & $111.2^{\tiny{+2.5}}_{\tiny{-1.4}} $& $8.60^{\tiny{+0.22}}_{\tiny{-0.89}} $& $19.00^{\tiny{+3.43}}_{\tiny{-2.57}}$ & 50.77 & True\\[2.5pt]  
$-560 $& $-100$ & $71.9^{\tiny{+1.5}}_{\tiny{-0.1}}$ & $5.30^{\tiny{+1.17}}_{\tiny{-0.54}} $& $11.10^{\tiny{+0.75}}_{\tiny{-6.22}}$ & 51.19 & True\\[2.5pt]  
$-680$ & $-190 $& $96.4^{\tiny{+10.9}}_{\tiny{-35.0}}$ & $6.60^{\tiny{+0.96}}_{\tiny{-0.49}}$ & $13.80^{\tiny{+2.26}}_{\tiny{-0.44}} $& 51.02 & True\\[2.5pt]  
$-530$ & $-190$ & $77.9^{\tiny{+11.3}}_{\tiny{-11.3}}$ & $5.30^{\tiny{+1.07}}_{\tiny{-0.42}}$ & $17.80^{\tiny{+2.76}}_{\tiny{-1.01}}$ & 51.13 & True\\[2.5pt]  
$-630$ & $-280$ & $88.8^{\tiny{+3.3}}_{\tiny{-0.3}}$ & $6.40^{\tiny{+0.49}}_{\tiny{-0.09}}$ & $20.70^{\tiny{+2.21}}_{\tiny{-0.17}}$ & 50.89 & True\\[2.5pt]  
$-850 $&$-100$ & $102.0^{\tiny{+5.7}}_{\tiny{-3.5}}$ & $8.20^{\tiny{+0.16}}_{\tiny{-0.61}}$ & $12.20^{\tiny{+0.57}}_{\tiny{-0.32}}$ & 50.95& False\\[2.5pt]  
$-270$ & $-360$ & $35.0^{\tiny{+0.4}}_{\tiny{-4.1}}$ & $3.50^{\tiny{+1.04}}_{\tiny{-0.06}} $& $40.10^{\tiny{+17.14}}_{\tiny{-0.82}}$ & 51.46 & False\\[2.5pt]  
$-370$ & $-270$ & $56.0^{\tiny{+1.3}}_{\tiny{-3.6}}$ & $4.30^{\tiny{+0.54}}_{\tiny{-0.32}}$ & $30.70^{\tiny{+12.04}}_{\tiny{-2.55}}$ & 51.43 & True\\[2.5pt]  
$-800 $& $-50$ & $96.5^{\tiny{+7.2}}_{\tiny{-7.9}}$ & $7.50^{\tiny{+0.48}}_{\tiny{-0.53}} $& $12.80^{\tiny{+0.83}}_{\tiny{-1.01}} $ & 51.01 & True\\[2.5pt]  
$-700$ & $-50$ & $86.6^{\tiny{+0.2}}_{\tiny{-0.4}}$ & $6.70^{\tiny{+0.49}}_{\tiny{-0.58}}$ & $13.90^{\tiny{+1.06}}_{\tiny{-1.14}}$ & 51.13 & False\\[2.5pt]  
$-500$ & $-50$ & $63.0^{\tiny{+10.3}}_{\tiny{-10.1}}$ & $4.90^{\tiny{+0.62}}_{\tiny{-0.55}}$ & $17.90^{\tiny{+1.81}}_{\tiny{-1.46}}$ & 51.30 & True\\[2.5pt]  
$-600$ & $-50$ & $75.5^{\tiny{+3.2}}_{\tiny{-3.0}} $ & $5.80^{\tiny{+0.21}}_{\tiny{-0.17}} $& $15.10^{\tiny{+0.18}}_{\tiny{-3.09}}$ & 51.23 & True\\[2.5pt]  
$-200$ & 100 & $22.0^{\tiny{+13.3}}_{\tiny{-12.9}} $ & $2.10^{\tiny{+0.68}}_{\tiny{-0.62}} $& $-25.00^{\tiny{+17.56}}_{\tiny{-17.54}}$ & 51.54 & False\\[2.5pt]  
200 & 100 & $23.0^{\tiny{+13.8}}_{\tiny{-14.0}}$ & $1.70^{\tiny{+0.61}}_{\tiny{-0.68}}$ & $-151.00^{\tiny{+22.29}}_{\tiny{-22.50}}$ & 51.48 & False\\[2.5pt]  
$-100$ & 250 & $30.0^{\tiny{+12.8}}_{\tiny{-12.8}}$ & $2.50^{\tiny{+0.61}}_{\tiny{-0.66}}$ & $-65.00^{\tiny{+14.22}}_{\tiny{-15.19}}$ & 51.48 & False\\[2.5pt]  
$-600$ & 400 & $90.6^{\tiny{+6.6}}_{\tiny{-6.6}}$ & $6.50^{\tiny{+0.62}}_{\tiny{-0.29}}$ & $-21.00^{\tiny{+0.36}}_{\tiny{-0.80}}$ & 50.97 & True\\[2.5pt]  
$-700 $ & 250 & $94.0^{\tiny{+5.1}}_{\tiny{-6.0}}$ & $7.10^{\tiny{+0.40}}_{\tiny{-0.33}} $& $-23.00^{\tiny{+2.84}}_{\tiny{-2.98}} $ & 51.08 & False\\[2.5pt]  
$-500 $& 250 & $74.0^{\tiny{+9.3}}_{\tiny{-10.1}}$ & $5.60^{\tiny{+0.54}}_{\tiny{-0.61}}$ & $-29.00^{\tiny{+6.30}}_{\tiny{-5.61}}$ & 51.25 & False\\[2.5pt]  
$-750$ & 0 & $91.3^{\tiny{+6.4}}_{\tiny{-6.4}}$ & $7.00^{\tiny{+0.58}}_{\tiny{-0.30}}$ & $0.00^{\tiny{+3.37}}_{\tiny{-3.37}}$ & 51.10 & True\\[2.5pt]  
$-800$ & $-600 $& $80.0^{\tiny{+9.9}}_{\tiny{-9.6}} $& $8.70^{\tiny{+0.68}}_{\tiny{-0.63}}$ & $31.00^{\tiny{+4.46}}_{\tiny{-4.76}}$ & 50.81 & False\\[2.5pt]  
$-600$  & $-600$ & $61.0^{\tiny{+3.5}}_{\tiny{-4.0}}$ & $8.00^{\tiny{+0.06}}_{\tiny{-1.15}}$ & $30.00^{\tiny{+1.85}}_{\tiny{-0.22}} $& 51.06 & False\\[2.5pt]  
$-450$ & 100 & $ 59.0^{\tiny{+6.3}}_{\tiny{-6.8}} $& $4.70^{\tiny{+0.34}}_{\tiny{-0.40}} $& $-13.00^{\tiny{+4.54}}_{\tiny{-4.53}} $& 51.35 & False\\[2.5pt]  
$-650$ & $-350$ & $93.5^{\tiny{+3.6}}_{\tiny{-3.6}} $& $7.00^{\tiny{+0.20}}_{\tiny{-0.28}}$ & $20.00^{\tiny{+1.33}}_{\tiny{-0.14}} $& 51.01 & True\\[2.5pt]  
$-650$ & $-100$ &$  82.8^{\tiny{+9.9}}_{\tiny{-9.8}} $& $6.39^{\tiny{+0.62}}_{\tiny{-0.62}} $& $10.00^{\tiny{+5.47}}_{\tiny{-5.63}}$ & 51.19 & False\\[2.5pt]  
$-580$ & 0 & $72.1^{\tiny{+4.1}}_{\tiny{-1.5}} $& $5.00^{\tiny{+0.38}}_{\tiny{-0.12}} $& $0.00^{\tiny{+4.49}}_{\tiny{-4.49}} $& 51.12 & True\\[2.5pt]  
$-850 $ & 0 & $96.6^{\tiny{+0.5}}_{\tiny{-0.6}} $& $7.00^{\tiny{+0.13}}_{\tiny{-0.17}} $& $0.00^{\tiny{+4.34}}_{\tiny{-4.34}} $& 50.95 & True\\[2.5pt]  
$-750 $& $-100$ & $93.0^{\tiny{+3.9}}_{\tiny{-3.9}} $& $7.21^{\tiny{+0.27}}_{\tiny{-0.26}}$ & $9.10^{\tiny{+1.97}}_{\tiny{-2.15}}$ & 51.09 & False\\[2.5pt]  
$-350$ & $-350$ & $26.3^{\tiny{+0.2}}_{\tiny{-2.0}}$ & $5.00^{\tiny{+0.31}}_{\tiny{-0.04}}$ & $47.00^{\tiny{+7.50}}_{\tiny{-11.30}}$ & $51.17 $& False\\[2.5pt]  
    \hline
    \end{tabular}
\caption{\textcolor{black}{We present in this table some of the calibrated horn positions and angles obtained using TICRA-GRASP. The True/False column indicates if the result was a primary or a secondary peak of the response function of the telescope for those locations of x and y. Error values were obtained only for variable parameters and within 0.01 dB.}}
\label{tab:general_table}
\end{table*}

\section{Horn Attenuation Angle
\label{AppendixB}}
\begin{figure*}
    \centering
    \includegraphics[scale = 0.22]{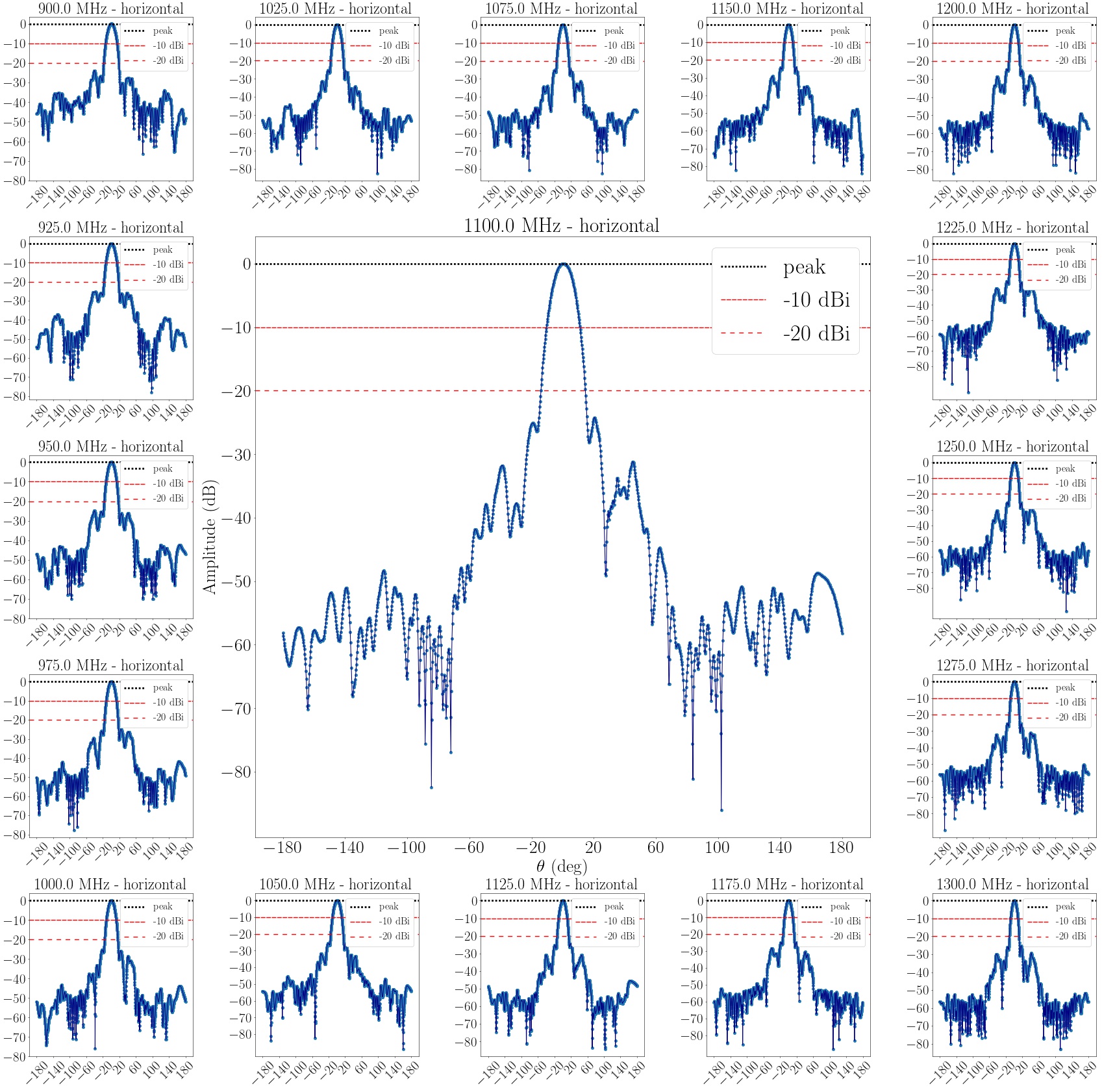}
    \caption{Plots from the horn beam measurements using data from \cite{2020ExA....50..125W}, They were used to obtain the 10-20 dB attenuation angles as a function of frequency. Results shown here are for a horizontally polarized input signal. For which plot the intensity is attenuated by 10\,dB (red dashed line) and 20\,dB (sparse red dashed line) in relation to the peak intensity (black dashed line).}
    \label{fig:LIT2008_h}
\end{figure*}

\begin{figure*}
    \centering
    \includegraphics[scale = 0.22]{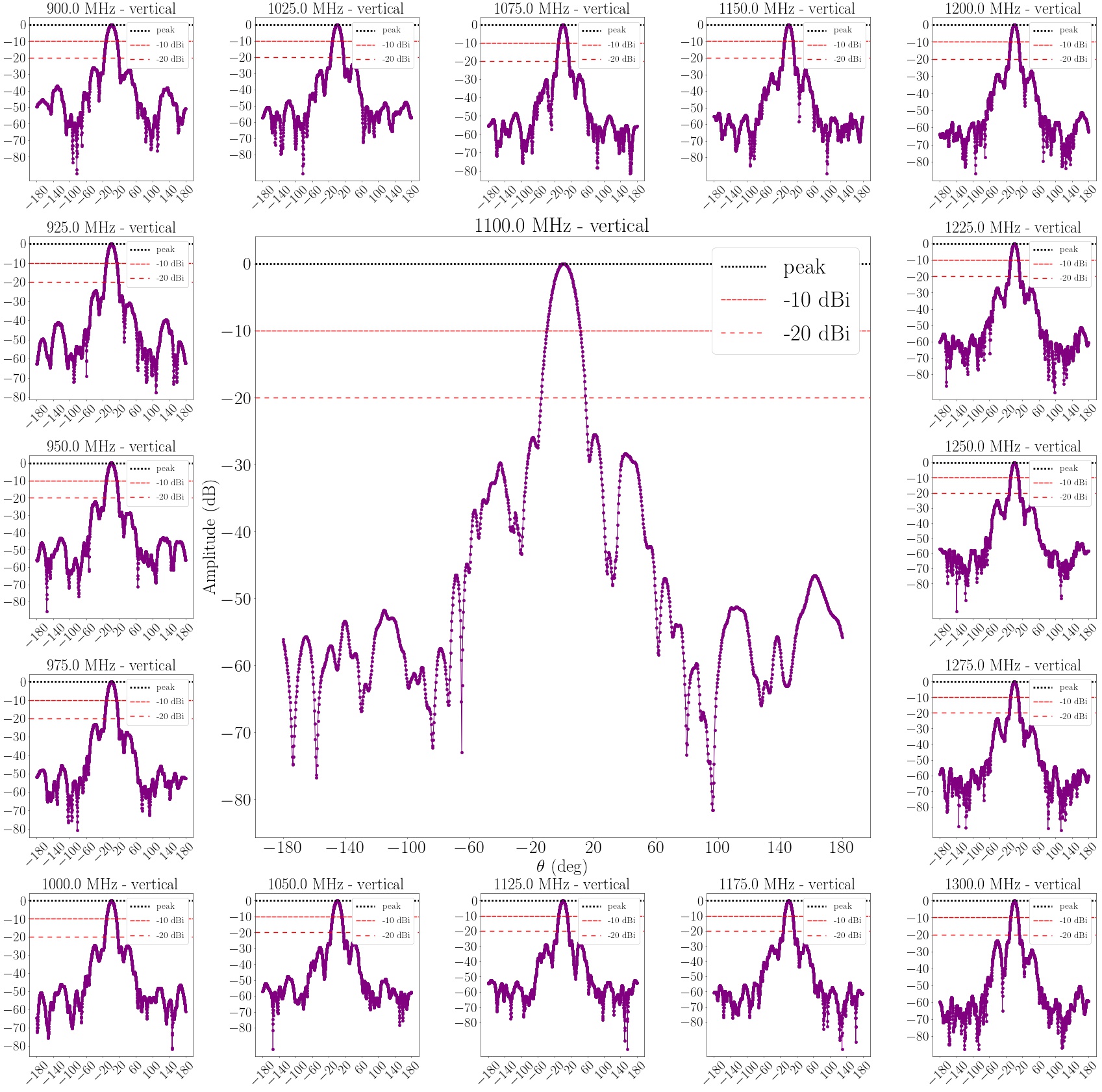}
    \caption{Same as Fig. \ref{fig:LIT2008_h}, but for a vertically polarized input signal. For which plot the intensity is attenuated by 10\,dB (red dashed line) and 20\,dB (sparse red dashed line) in relation to the peak intensity (black dashed line).}
    \label{fig:LIT2008_v}
\end{figure*}

We interpolate the beam data (horn response) and built two sets of data, for each beam component and each frequency, with intensity decrease value of 10 and 20\,dB in relation to maximum intensity value (first lobe) and we take the corresponding angle values, as can seen in Fig. \ref{fig:LIT2008_h} for horizontal component of the beam and in Fig. \ref{fig:LIT2008_v} for vertical one.  In Fig. \ref{fig:LIT2008_h} we can see the main results separated by correlation between maximum and minimum angles, frequency and amplitude of the intensity. First and third plots in first row are relations between minimum and maximum angles with its amplitude value for horizontal and vertical component, respectively. Salmon and red colors are related to decrease in intensity of 10\,dB for minimum and maximum angle, respectively, and light blue and navy blue for 20\,dB. Each color represent a group and is basically nested in the same region, but there is a more distant triangle for each group that are values from 900\,MHz, where we have a first lobule with greater amplitude and wider (see  Figures \ref{fig:LIT2008_h} and  \ref{fig:LIT2008_v}). Those can also be seen in the plots between frequency and angle (second and fourth plots in first row). We can also see in those plots that the higher frequency, the lower the angles that delimit the peak interval. In the second row, first and third plots relate frequency and absolute angle, for minimum and maximum angle value, and its mean angle. The absolute value decreases with increasing frequency. Second and fourth plots are the mean weighted values given by \begin{equation}
\theta_{\rm weighted}(\nu) = \frac{\nu\ \textrm{(MHz)}}{1100\ \textrm{(MHz)}}\theta (\nu)\,,
\end{equation}
where the decreases are smooth. 
Using a range between minimum and maximum angles we fitted the Gaussian function to the peak in order to estimate the 10-20 dB attenuation angles for the BINGO horn as a function of frequency. The results are fitted with the following function
\begin{equation}
G(\theta|a,b,c,d) = ae^{-0.5\frac{(\theta-b)^2}{c^2}}+d,
\label{taper_angle_fit}
\end{equation}
for both, 10 and 20\,dB. We center the Gaussian beam in  $b= 0^{\circ}$ and take the new angles for 10 and 20\,dB intensity decreases and each frequency. The results for 10\,dB are shown in Fig. \ref{fig:LIT2008_interpolation}. Then, the attenuation angles were obtained by fitting
\begin{equation}
\theta_{\textrm{taper}} (\nu |a,b,c,d,f)= a + b\nu + c\nu^2 + \frac{d}{\nu} + \frac{f}{\nu^2}\,.
\end{equation}

\newpage
\mbox{~}
\clearpage

\end{appendix}

\end{document}